\documentclass[aps,superscriptaddress,twocolumn,prl]{revtex4-1}
\usepackage{amssymb,amsmath,mathptmx}
\usepackage{graphicx}
\usepackage{dcolumn}
\usepackage{bm}
\usepackage{hyperref}
\usepackage{color}
\usepackage[utf8x]{inputenc}
\usepackage{array}
\usepackage{ulem}
\newcolumntype{C}{>{\centering\arraybackslash}p{1em}}

\newcommand{\be}{\begin{equation}}
\newcommand{\ee}{\end{equation}}
\newcommand{\bea}{\begin{eqnarray}}
\newcommand{\eea}{\end{eqnarray}}

\begin{document}

\title{Floquet-Multiple Andreev Reflections}

\author{R\'egis M\'elin}
\email{regis.melin@neel.cnrs.fr}

\affiliation{Universit\'e Grenoble-Alpes, CNRS, Grenoble INP, Institut
 NEEL, Grenoble, France}

\author{Romain Danneau}

\affiliation{Institute for Quantum Materials and Technologies,
 Karlsruhe Institute of Technology, Karlsruhe D-76021, Germany}

\author{Morteza Kayyalha}

\affiliation{Department of Electrical Engineering, The Pennsylvania
State University, University Park, Pennsylvania 16802, USA}

\begin{abstract}
  Floquet theory describes quantum systems governed by time-periodic
  Hamiltonians, much as Bloch theory describes spatially periodic
  solids. In voltage-biased multiterminal Josephson junctions, the
  Josephson relation causes superconducting phase differences to
  evolve periodically in time, thereby providing an intrinsic Floquet
  drive. In this Letter, we consider three-terminal Josephson
  junctions formed on a ballistic two-dimensional normal conductor
  with a continuum of electronic states. We show that the quartet and
  higher-order multipair processes yield characteristic
  Floquet-multiple Andreev reflection (Floquet-MAR) finite-bias
  conductance and noise resonances that are parameterized by the bias
  voltage and electrochemical potential. This microscopic picture
  opens a route toward implementing and probing Floquet-MAR physics in
  ballistic multiterminal Josephson junctions.
\end{abstract}
\maketitle

The physics of time periodic phenomena can be described by the Floquet
formalism \cite{Floquet1883}. In quantum systems, an external periodic
drive allows the creation and control of nonequilibrium Floquet states
such as in topological materials \cite{Oka2019}, ultracold atoms
\cite{Holthaus2019} as well as Floquet qubits
\cite{Gandon2022,Nguyen2024}. For example, a voltage-biased
two-terminal Josephson junction driven by an ac excitation gives rise
to discrete Floquet resonances \cite{Park2022}. Such spectra could
also be generated in superconducting multiterminals while the extra
leads play the role of the external periodic drive
\cite{Cuevas-Pothier,Freyn2011,Melin2017,Melin2019,Melin2020,Keliri2023a,Keliri2023b}.

Among the striking nonequilibrium effects in
  superconducting weak links \cite{Likharev}, multiple Andreev reflections (MARs)
  describe coherent quasiparticle transport involving repeated Andreev
  processes and multipair charge transfer across a Josephson
  junction \cite{Octavio1983,Averin1995,Cuevas1996}. While MAR physics has been studied extensively in both
  two-terminal and multiterminal settings \cite{Octavio1983,Averin1995,Cuevas1996,Duhot2009,Houzet2010,Jonckheere,Gosselin2014,Melin2016,Kraft2018,Draelos2019,Pankratova2020,Arnault2021,Coraiola2023,Keliri2023a,Keliri2023b,Jung2025}, its interplay with Floquet
  physics in multiterminal devices has remained largely
  unexplored. Here, we develop a Floquet-based transport description
  for two-dimensional multiterminal Josephson junctions, where the
  electrochemical potential $\mu_N$ of the normal region serves as an
  independent spectroscopic control parameter in addition to the bias
  voltage $V$ and magnetic flux $\Phi$. Using Keldysh Green’s
  functions, we calculate the current susceptibility $\chi_I =
  \partial I(eV,\mu_N)/\partial \mu_N$ and identify resonances that we
  interpret as Floquet processes dressed by multiple Andreev
  reflections, which we call Floquet-MARs. More broadly, our results
  suggest that multiterminal Josephson junctions can provide a
  controllable platform for studying nonequilibrium Floquet-Andreev
  spectra.

The term {\it Floquet-MAR resonances} refers to the finite-bias
resonant features in the differential current and quantum noise
susceptibilities, arising from Floquet processes dressed by MAR-like
energy shifts. These broadened spectral structures are generated by
coherent multichannel propagation through the ballistic continuum,
rather than discrete bound-state levels. The higher-order dimer,
trimer, and quadruplet structures discussed below refer to
symmetry-related families of these resonances.

The Floquet-MAR finite-bias resonances in the current susceptibility
can be understood as arising from the underlying Andreev-tube
structure of the multichannel ballistic configurations. These Andreev
tubes correspond to families of quasi-one-dimensional semiclassical
trajectories formed by counterpropagating electron and
Andreev-reflected hole components,
see e.g. \cite{Kraft2018,Meier2016,Rashid2025,Rashid2026}.

{\it Devices and Hamiltonians:} Our framework combines three distinct
elements: a microscopic picture based on Andreev-tube trajectories, a
spectroscopic theory of Floquet-MAR resonances in nonlinear
conductance, and a quantum noise diagnostic of correlated multipair
transport. This separation between mechanism, observable signatures,
and noise fingerprints provides a clear physical interpretation of the
predicted Floquet-MAR phenomena.

\begin{figure}[htb]
  \centerline{\includegraphics[width=.575\columnwidth]{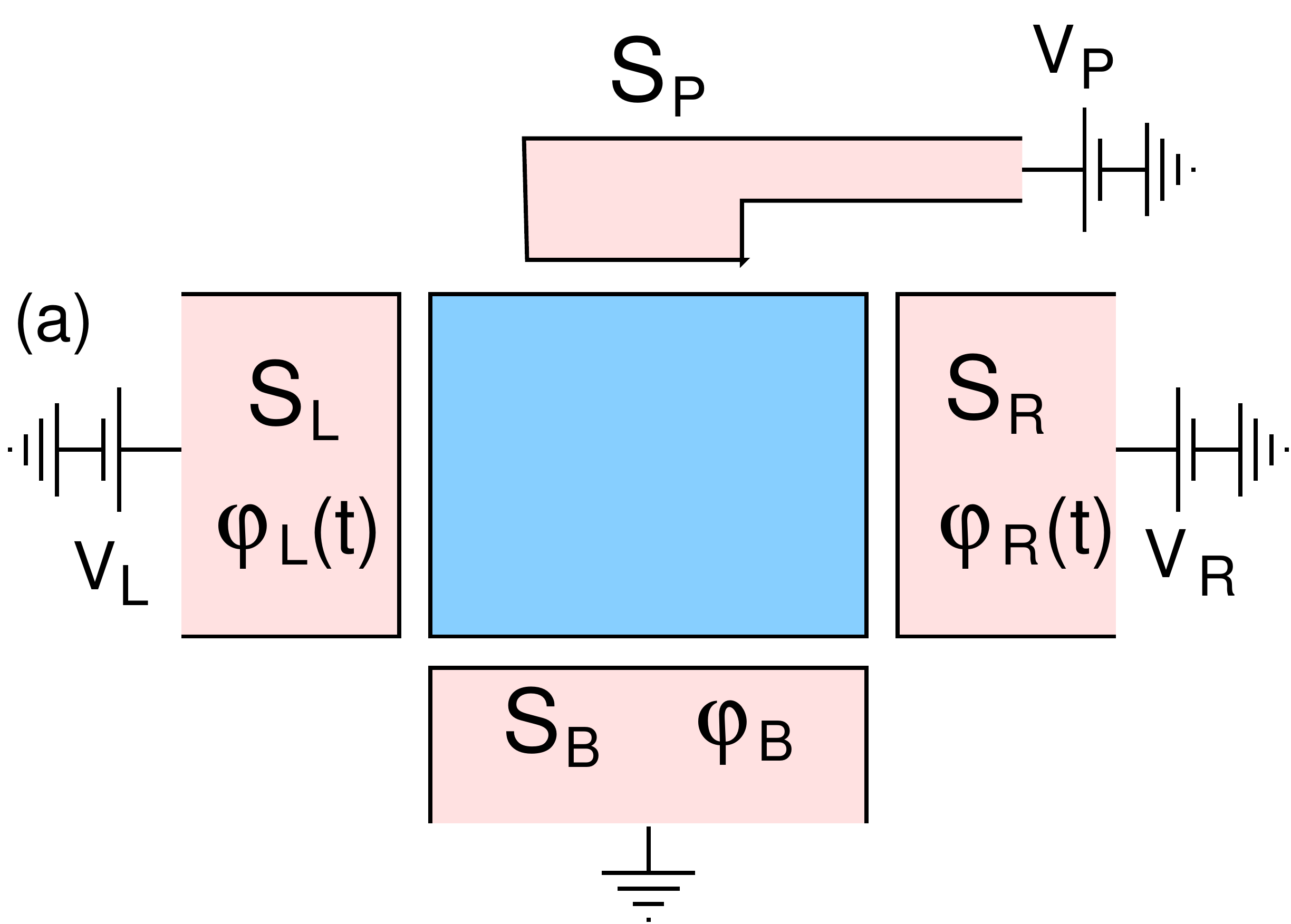}\includegraphics[width=.425\columnwidth]{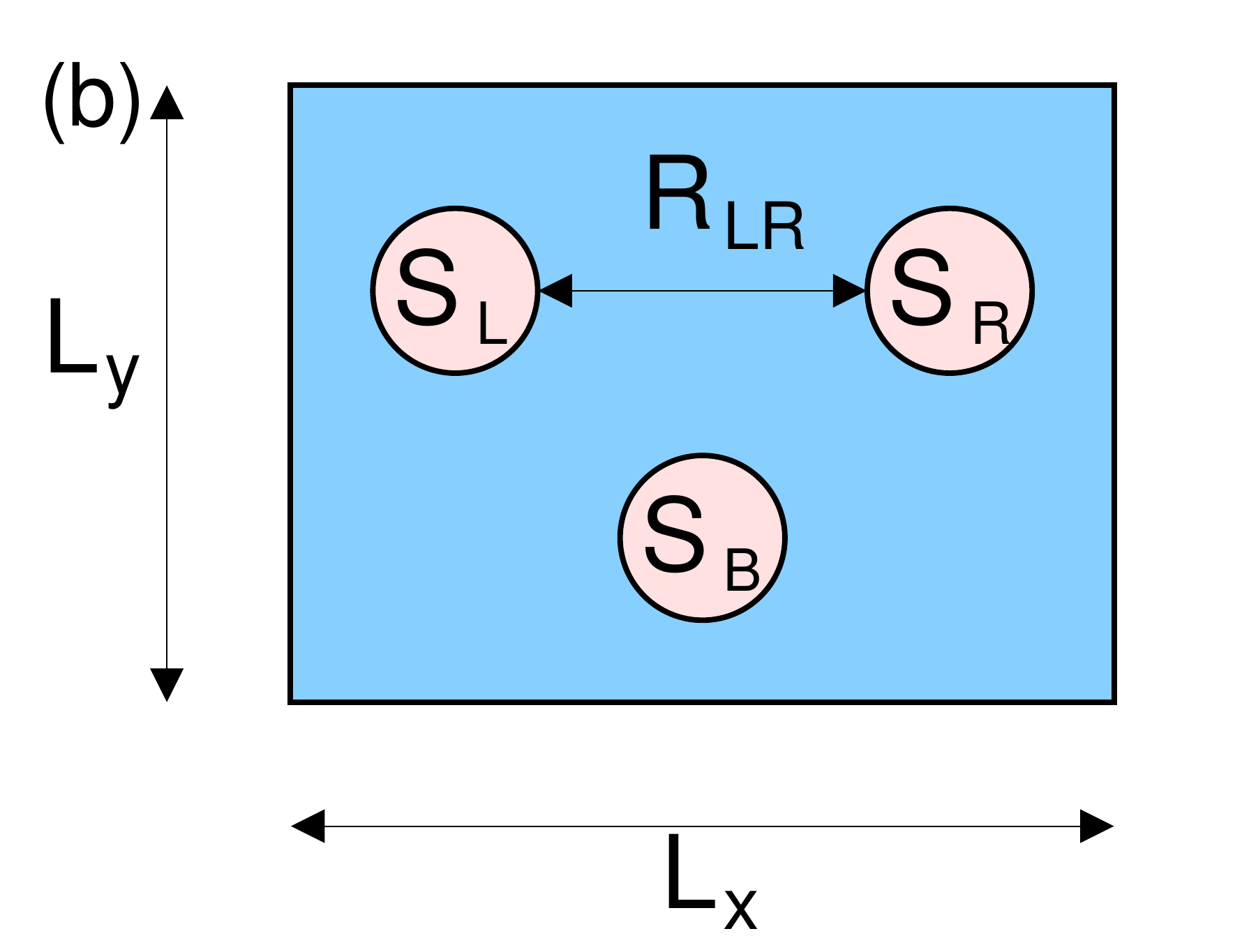}}

  \caption{{\it Set-up of our study:} The square-shape 2D
    conductor-based 3TJJ of dimension $L_x\times L_y$ (a), biased at
    the voltages $(V_L,\,V_R,\,V_B)$, with the superconducting phase
    variables $\varphi_L(t)=\varphi_L+2eV_L t/\hbar$,
    $\varphi_R(t)=\varphi_R+2eV_R t/\hbar$ and $\varphi_B$ at the time
    $t$. Top view of the model geometry (b), to obtain a continuum in
    the weak link, we set the dimensions of the 2D conductor $L_x$ and
    $L_y$ to infinity while the superconducting leads $S_L$, $S_R$ and
    $S_B$ are then model as points separated by $R_{k,l}$, where
    $k,\,l=L,\,R,\,B$. The superconducting leads therefore form
    multichannel point contacts on the 2D conductor, having the
    aperture $\pi d^2/4$ such that $d\gg\lambda_F$, with $\lambda_F$
    the Fermi wave-length. {We assume that an
      additional superconducting tunneling probe $S_P$ is connected to
      the 2D conductor (see panel a), so as to control the value of
      the electrochemical potential $\mu_N$
      \cite{Rashid2025,Rashid2026}.}
  \label{fig:1}}
\end{figure}

The voltage biasing is such that the superconductors $S_L$, $S_R$ and
$S_B$ are connected to a rectangular ballistic two-dimensional (2D)
conductor at $V_L$, $V_R$ and $V_B=0$, see Fig.~\ref{fig:1}a, with the
following time-evolution of the superconducting phase variables:
$\varphi_L(t) = \varphi_L + 2eV_Lt/\hbar$, $\varphi_R(t) = \varphi_R +
2eV_Rt/\hbar$ and $\varphi_B(t) \equiv \varphi_B$.

Concerning the geometry, we work in the large-system limit, where the
single-particle level spacing in the ballistic 2D conductor is
negligible compared to the energy scales relevant for transport,
allowing the normal region to be treated as having a continuum
spectrum. Thus, we theoretically implement the simplifying assumption
of an infinite ballistic 2D conductor geometry with
$L_x,\,L_y\rightarrow \infty$ in the top-view of Fig.~\ref{fig:1}b. On
this figure, the aperture of the point contact is such that the
diameter $d\gg\lambda_F$ is much larger than the Fermi wave-length
$\lambda_F$. Then, propagation between the contacts separated by the
$R_{k,l}$ (with $k,l=L,\,R,\,B$) is captured by standard microscopic
Green's functions calculations
\cite{Caroli1,Caroli2,Cuevas,Cuevas-noise}, where the wave-vectors
continuously vary as a function of the energy. The model also holds
for the large-scale interfaces, within summation over the conduction
channels.

The superconducting Hamiltonians take the standard BCS-form, see
section IA of the Supplemental Material (SM) \cite{supinfo}. The
ballistic 2D conductor is described by the generic tight-binding
Hamiltonians of free electrons on the hexagonal or square lattices
gated far from the midgap singularities, with the effective parameters
of the Fermi wave-vector $k_F$ and Fermi velocity $v_F$, see section
IB of the SM \cite{supinfo}.  {The superconducting leads and the 2D
  conductor are coupled by a standard interfacial hopping amplitude,
  see section IC of the SM \cite{supinfo}.}

The following calculations use a semiquantitative large-gap
approximation, which has proven useful in several contexts of
mesoscopic superconductivity
\cite{Zazunov2003,Meng2009,Melin2022,Klees2020}. This approximation
does not describe the full finite-gap MAR ladder of conventional
junctions. Rather, here Floquet-MAR denotes MAR-like energy shifts and
diagrammatic dressing of low-energy Floquet processes propagating
through the ballistic 2D continuum. At low bias and small interface
transparency, restoring a finite gap changes the quantitative energy
dependence and overall scale, but does not introduce additional
lowest-order diagrams. In this sense, the large-gap approximation
captures the low-energy Floquet-MAR resonance structure addressed
below.
  
\begin{figure}[htb]

  \begin{minipage}{.8\columnwidth}
 \begin{minipage}{\columnwidth}
  \begin{minipage}{.49\textwidth}
  \includegraphics[width=\textwidth]{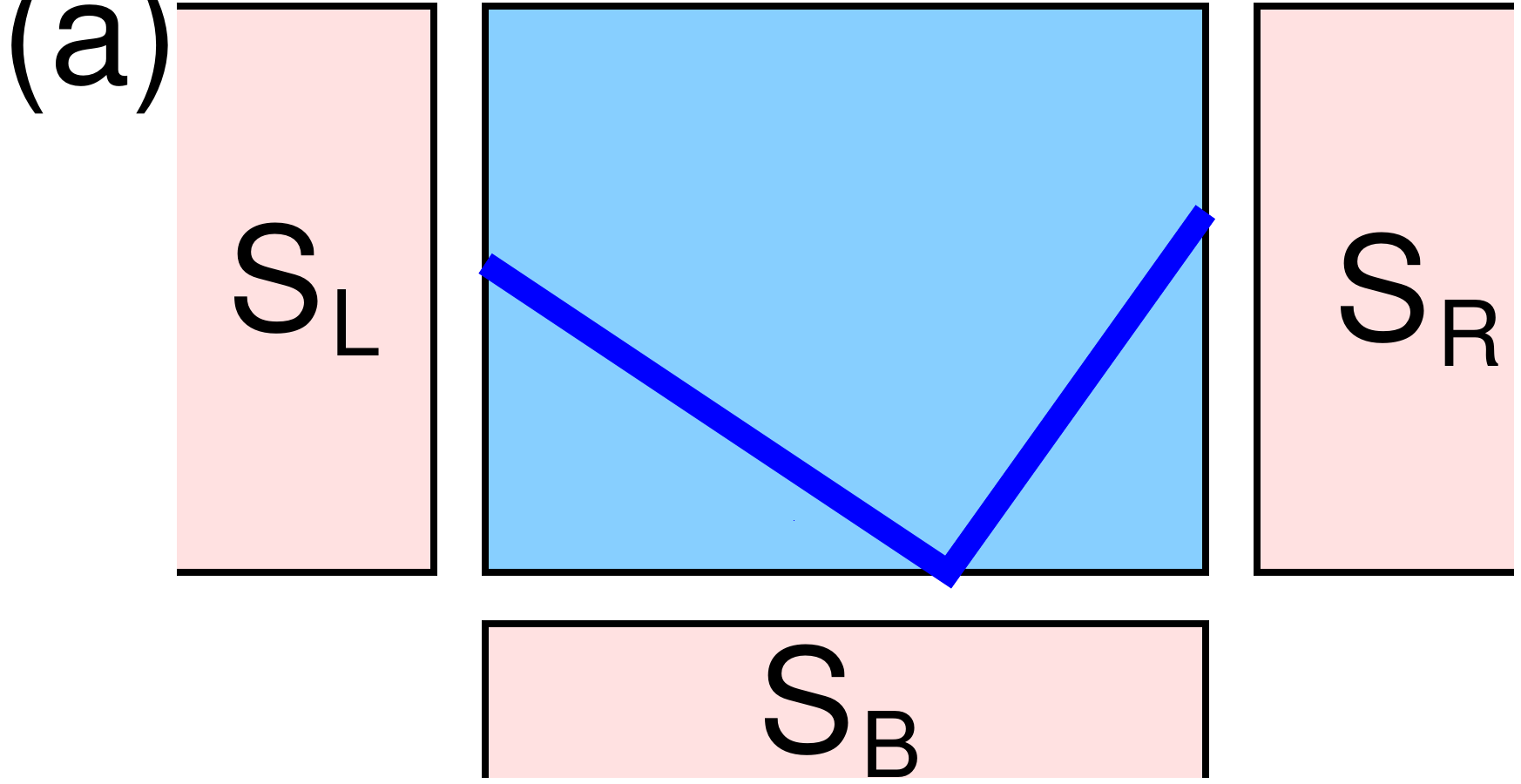}
  \end{minipage}
  \begin{minipage}{.49\textwidth}
  \hspace*{-1.1cm}\includegraphics[width=.55\textwidth]{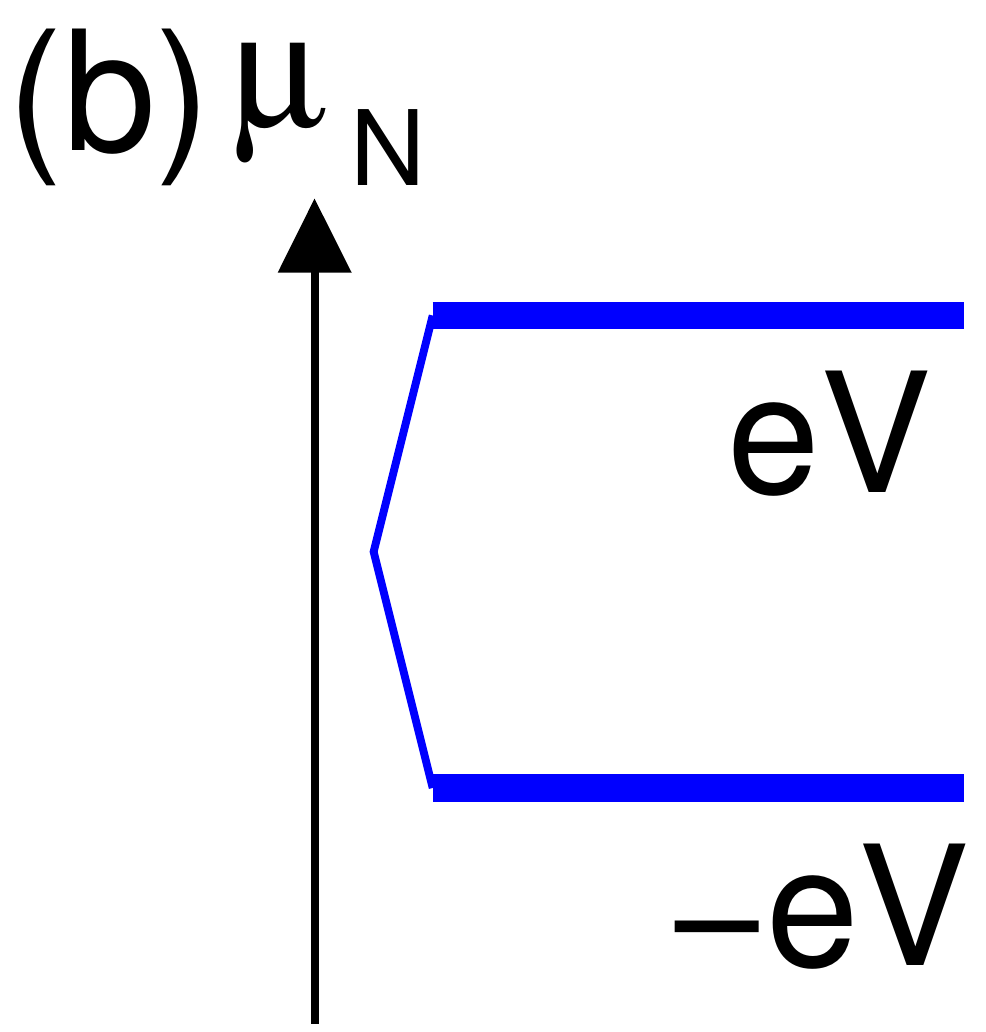}
  \end{minipage}
  \end{minipage}
  \begin{minipage}{\columnwidth}
  
    \begin{minipage}{\columnwidth}
      \begin{minipage}{.49\textwidth}
  \includegraphics[width=\textwidth]{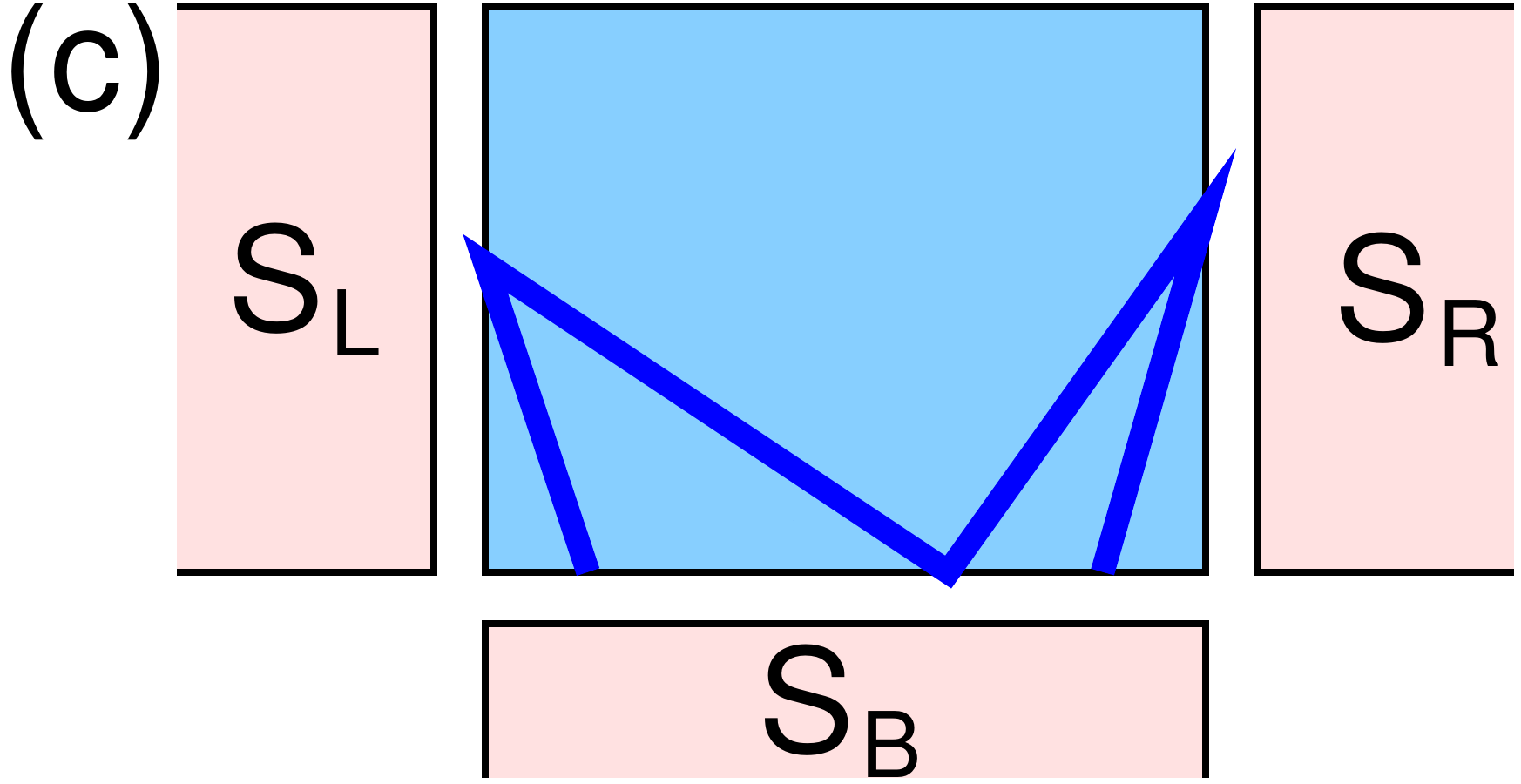}
    \end{minipage}
      \begin{minipage}{.49\textwidth}
  \hspace*{-.97cm}\includegraphics[width=.57\textwidth]{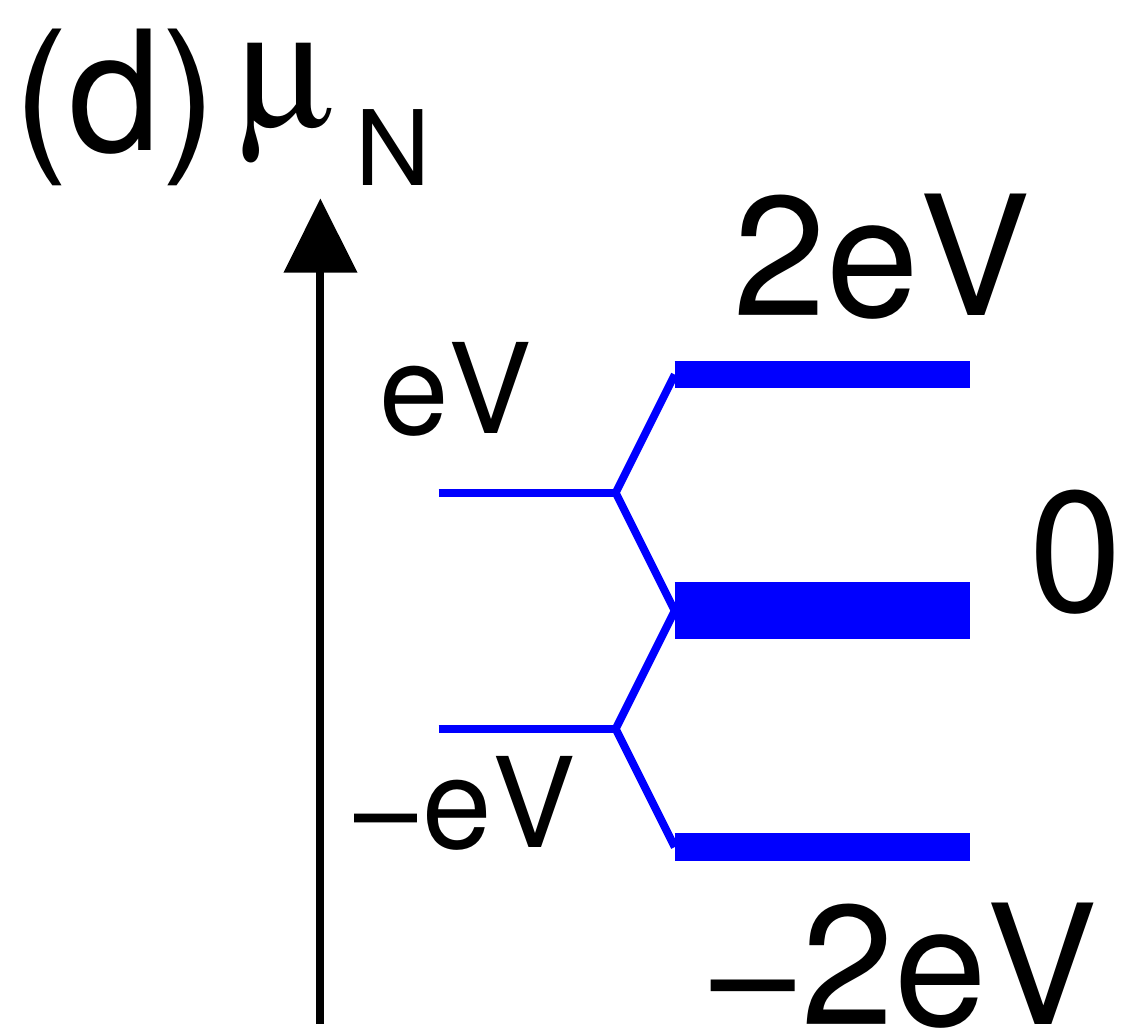}
  \end{minipage}
    \end{minipage}

  \begin{minipage}{\columnwidth}
  \begin{minipage}{.49\textwidth}
  \includegraphics[width=\columnwidth]{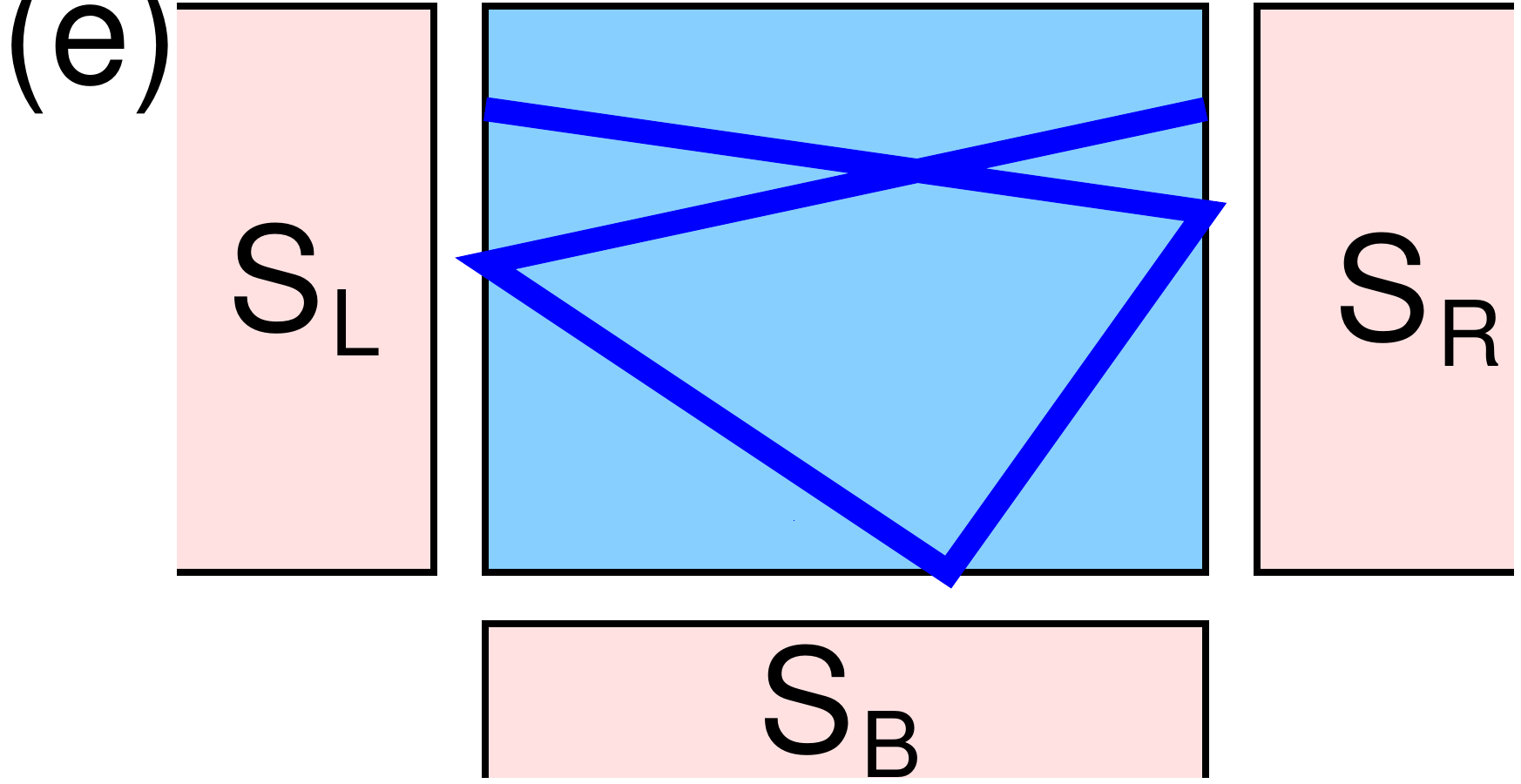}
    \end{minipage}
  \begin{minipage}{.49\textwidth}
  \includegraphics[width=.8\columnwidth]{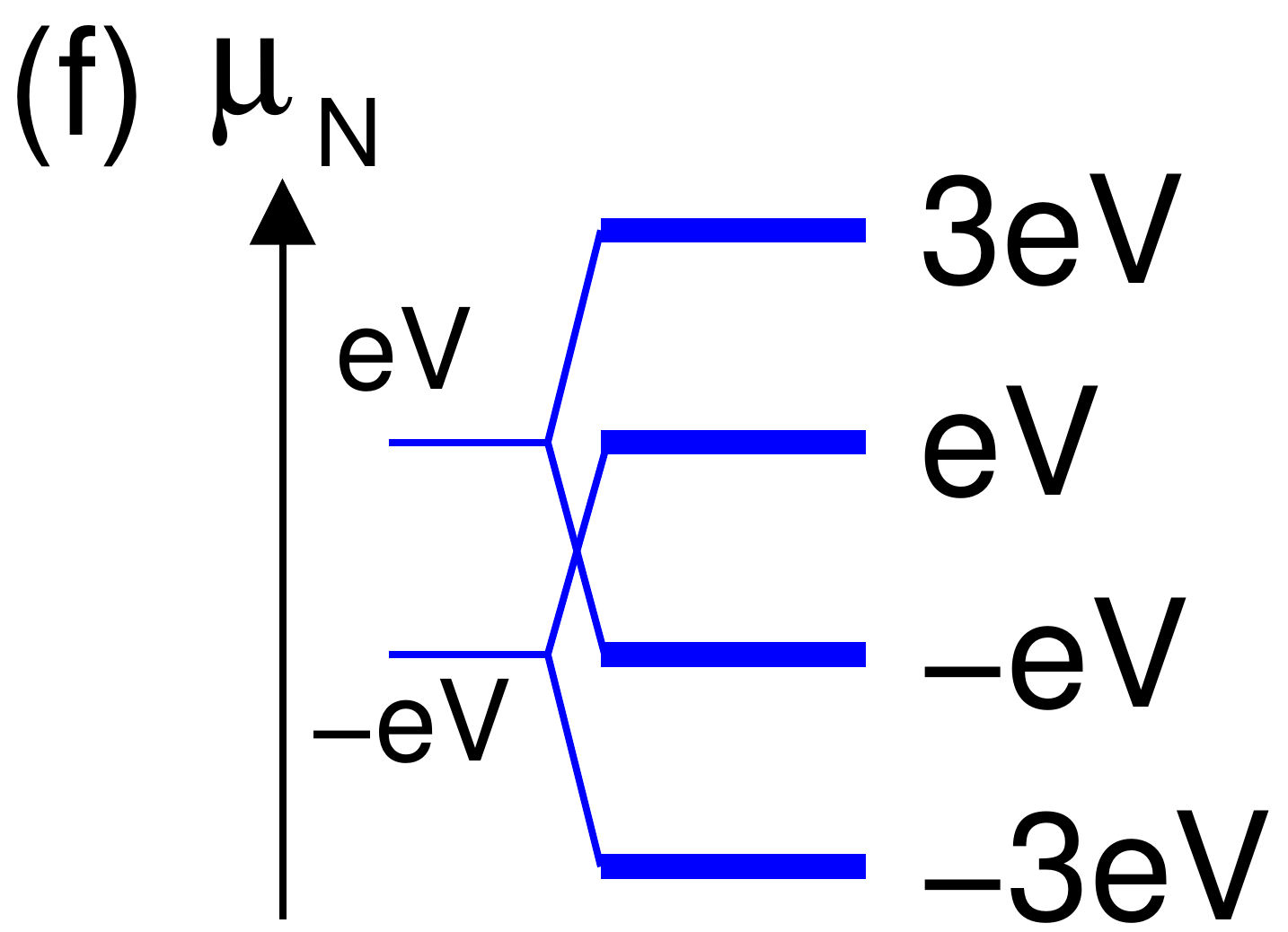}
      \end{minipage}
\end{minipage}
  \end{minipage}
  \end{minipage}
  \caption{ {\it Structures of the Andreev pair semiclassical
      trajectories:} (a), (c), (e) and {\it their corresponding energy
      level diagrams:} (b), (d) and (f). On panel (a), (c), and (e),
    the dark blue lines represent the nonlocal transmission of the
    Cooper pairs across the 2D conductor (a single line encodes an
    electron-electron e-e and hole-hole h-h Nambu Green's
    functions). (a) corresponds to a ${Q}$-quartet simplified
    semiclassical trajectories diagram, (c) an ${O}$-octet simplified
    semiclassical trajectories diagram and (e) a ${Q'}$-quartet
    simplified semiclassical trajectories diagram together with their
    corresponding energy level diagram, (b), (d) and (f) respectively.
  \label{fig:2}}
\end{figure}

\begin{figure}[htb]
  \begin{minipage}{.85\columnwidth}
  \centerline{\includegraphics[width=.8\columnwidth]{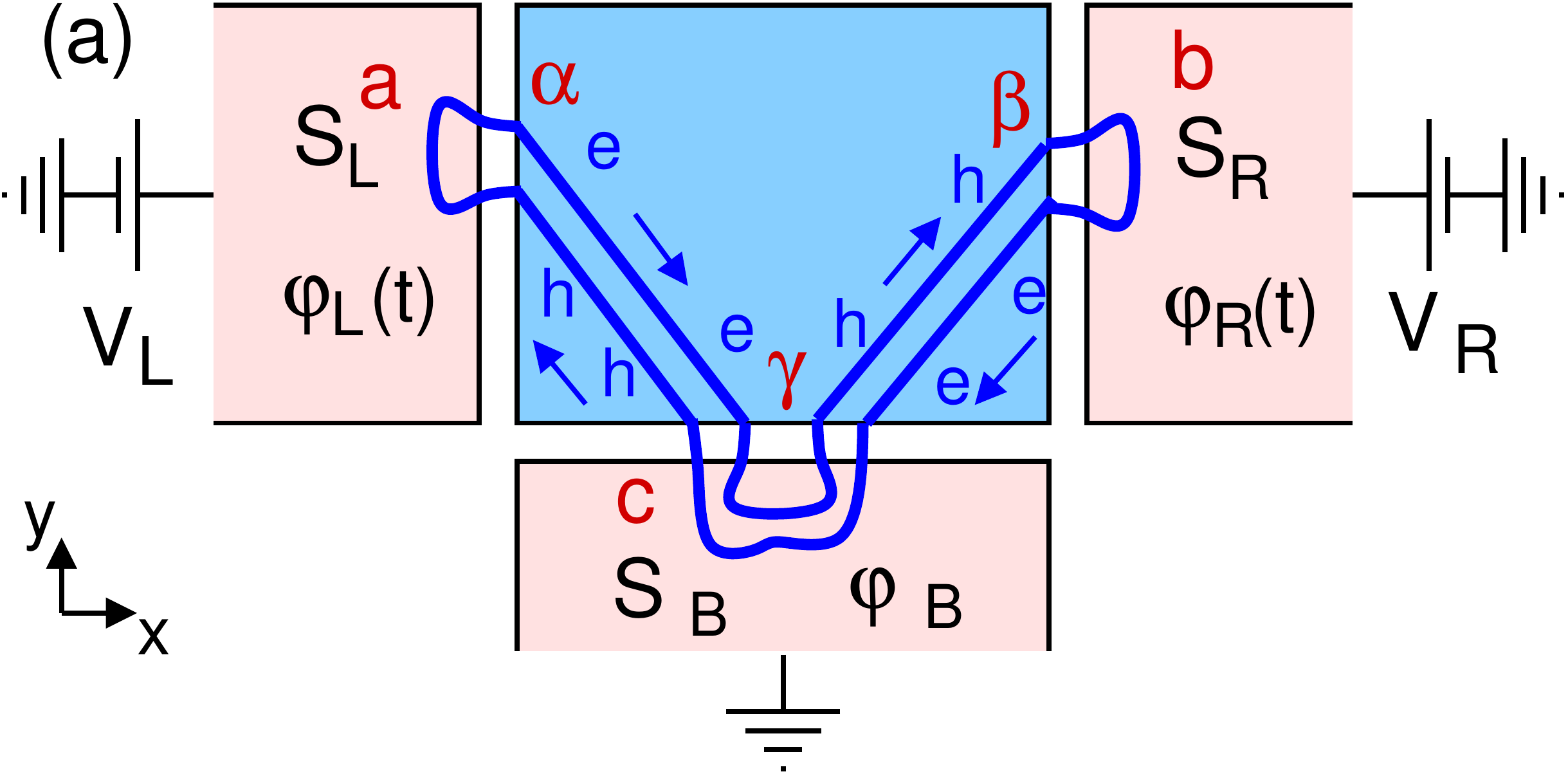}}
  \centerline{\includegraphics[width=.8\columnwidth]{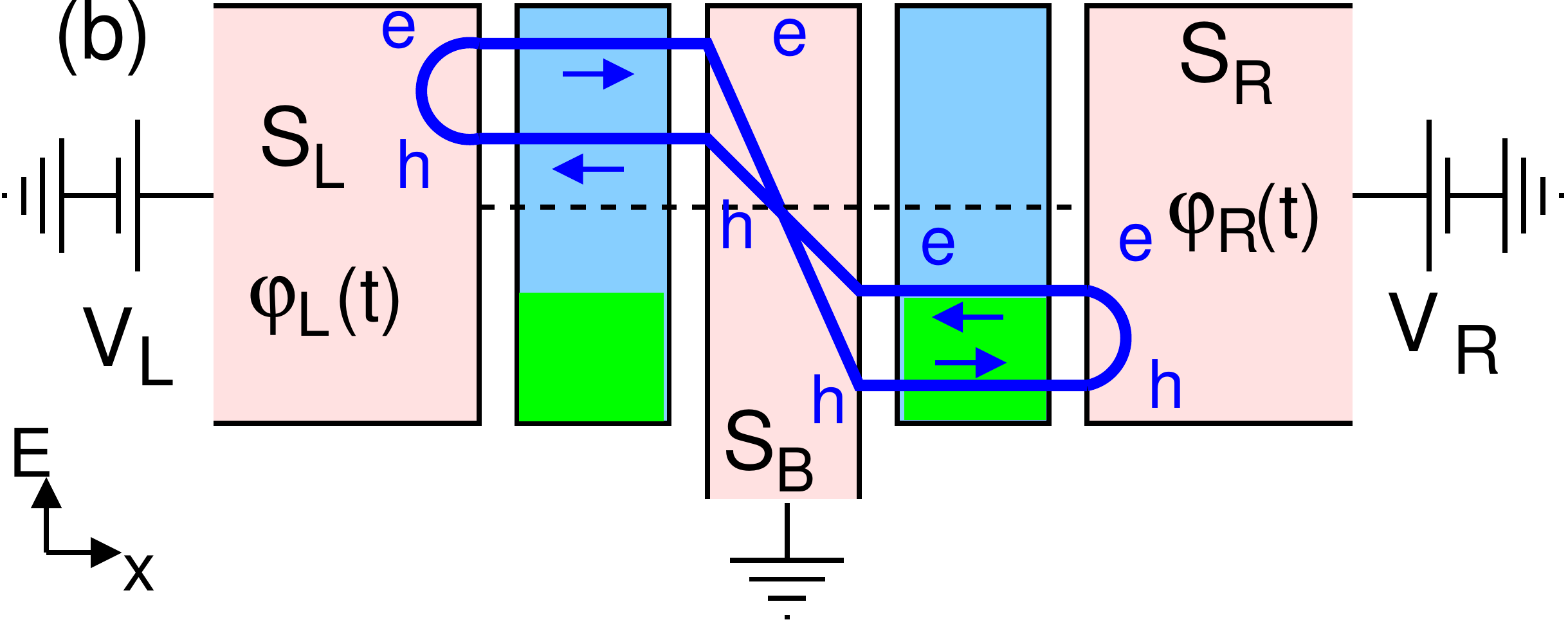}}
  \centerline{\includegraphics[width=.8\columnwidth]{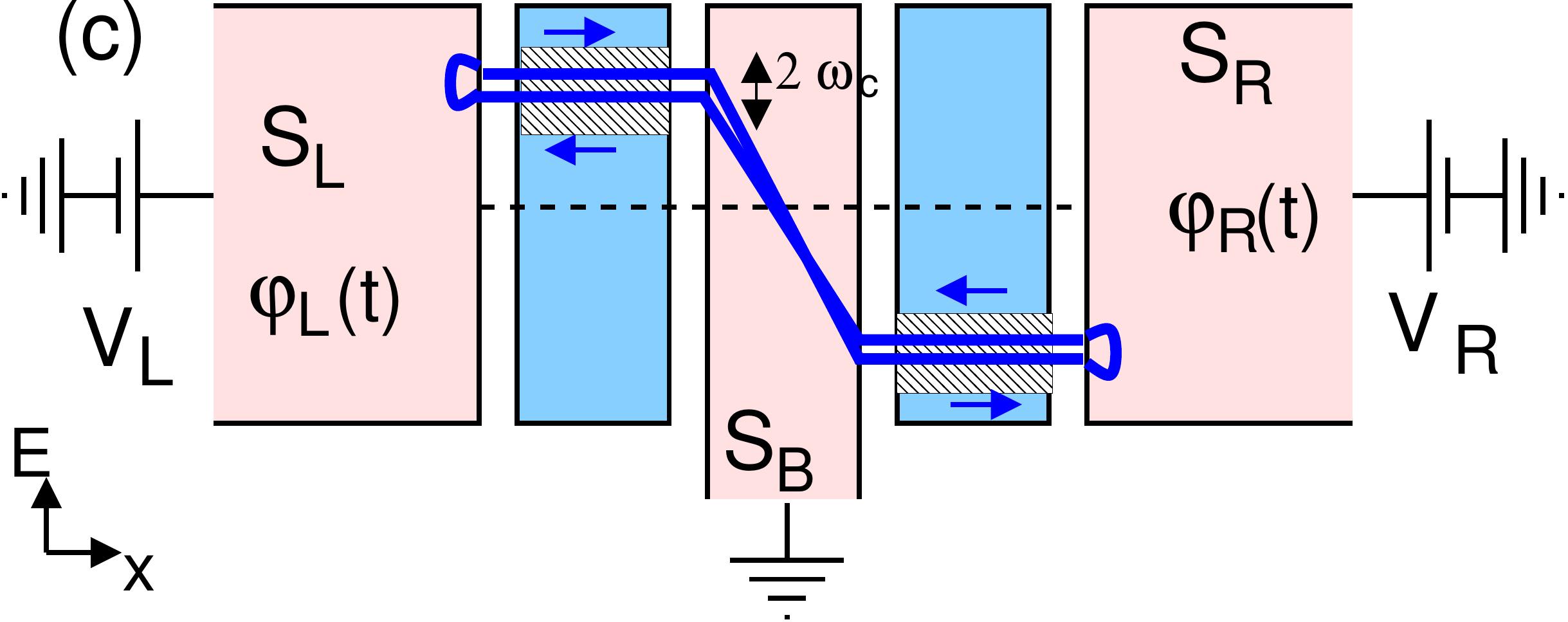}}
  \end{minipage}
  \caption{{\it The lowest-order ${Q}$-quartet diagram} in real space
    (a), and in the energy representation (b and c). The
    electrochemical potential fixes the discontinuity in the
    nonequilibrium Fermi surface at the energy $E=\mu_N$, see the
    representation of the filled states in green-color on panel
    b. {A finite-bias resonance in the current
      susceptibility} is obtained in transport if $E=\mu_N$ matches
    one of the intermediate-state Floquet-MAR energy levels of the
    quartets, as shown on panel b. Panel c shows a quasi-zero-energy
    quartet diagram restricted to $|\omega|\alt\omega_c$. The
    notations $e$ and $h$ are used for the electron-like and hole-like
    Nambu Green's functions, which are equivalently denoted by $1$ and
    $2$ in the analytical expressions. The notations $a$, $b$ label
    the quartet diagram tight-binding sites in the $S_L$ and $S_R$
    leads, and $\alpha$, $\beta$ are their counterpart in the central
    2D conductor. The notations $\gamma$ and $c$ are used for the
    contact with the bottom superconductor $S_B$.
    \label{fig:diagrammes}}
\end{figure}


The {\it quartets}
\cite{Pfeffer2014,Cohen2018,Huang2022,Graziano2022,Ohmacht2024,Gupta2024}
are the exotic multiterminal Josephson modes that can be probed at
equilibrium \cite{Ohmacht2024,Gupta2024} or at opposite bias voltages
\cite{Pfeffer2014,Cohen2018,Huang2022,Graziano2022}.  Multipairs such
as the sextets, octets and so on, are generally produced at
commensurate voltage biasing \cite{Jonckheere} or at equilibrium
\cite{sextets2025}.  The quartets microscopically originate from the
current-carrying exchange between the partners of the individual
Cooper pairs. {In a three-terminal junction, the
  elementary quartet process may be viewed as two Cooper pairs
  originating from the grounded lead $S_B$, which recombine coherently
  into one pair in $S_L$ and one pair in $S_R$. The net transferred
  charge is $4e$, and energy conservation requires
\[
2eV_L+2eV_R=0,
\]
yielding the quartet resonance condition $V_R=-V_L$.}  Different
approaches have developed to characterize the macroscopic signatures
of this multiparty Cooper pair entanglement, including conductance
maps as a function of the bias voltages
\cite{Pfeffer2014,Cohen2018,Huang2022,Gupta2024,Ohmacht2024}, quantum
noise \cite{Melin-Sotto}, equilibrium measurements \cite{Ohmacht2024}
such as tunneling spectroscopy of a bisquid \cite{Rech} or reentrant
critical current contours \cite{Melin2023a,Gupta2024}. We here propose
to resolve the energy spectrum of the quartet intermediate state in
large-scale ballistic 2D conductor-based MJJs, from the point of view
of the nonequilibrium current susceptibility spectroscopy of the
corresponding Floquet-MAR multiplets, see Figs.~\ref{fig:2}
and~\ref{fig:diagrammes}.

{\it Floquet-MARs:} Coming back to the 0D multiterminal quantum dots,
an approximation to the Floquet levels can be obtained from the
semiclassical Bohr-Sommerfeld quantization of the energy over the
running phase
\cite{Melin2017,Melin2019,Melin2020a,Doucot2020,Keliri2023a,Keliri2023b}.
{A useful interpretation of these spectra is to regard
  elementary Floquet {finite-bias resonances in the
    conductance} as being replicated and dressed by MAR-like
  processes. In contrast to conventional MAR, which connects subgap
  states to continuum quasiparticle states outside the superconducting
  gap, the present Floquet-MAR processes remain confined to the
  low-energy sector and generate characteristic resonance structures.}
Taking into account the presence/absence of zero-energy crossings in
the superconducting dispersion relations, Figs.~4a, b and c in
Ref.~\onlinecite{Melin2019} can be interpreted as the Floquet-MAR
replica of the bare quantum dot zero-energy level, averaged over one
period of oscillations.
\begin{figure*}[htb]
  \begin{minipage}{.9\textwidth}
  \centerline{\includegraphics[width=.3\textwidth]{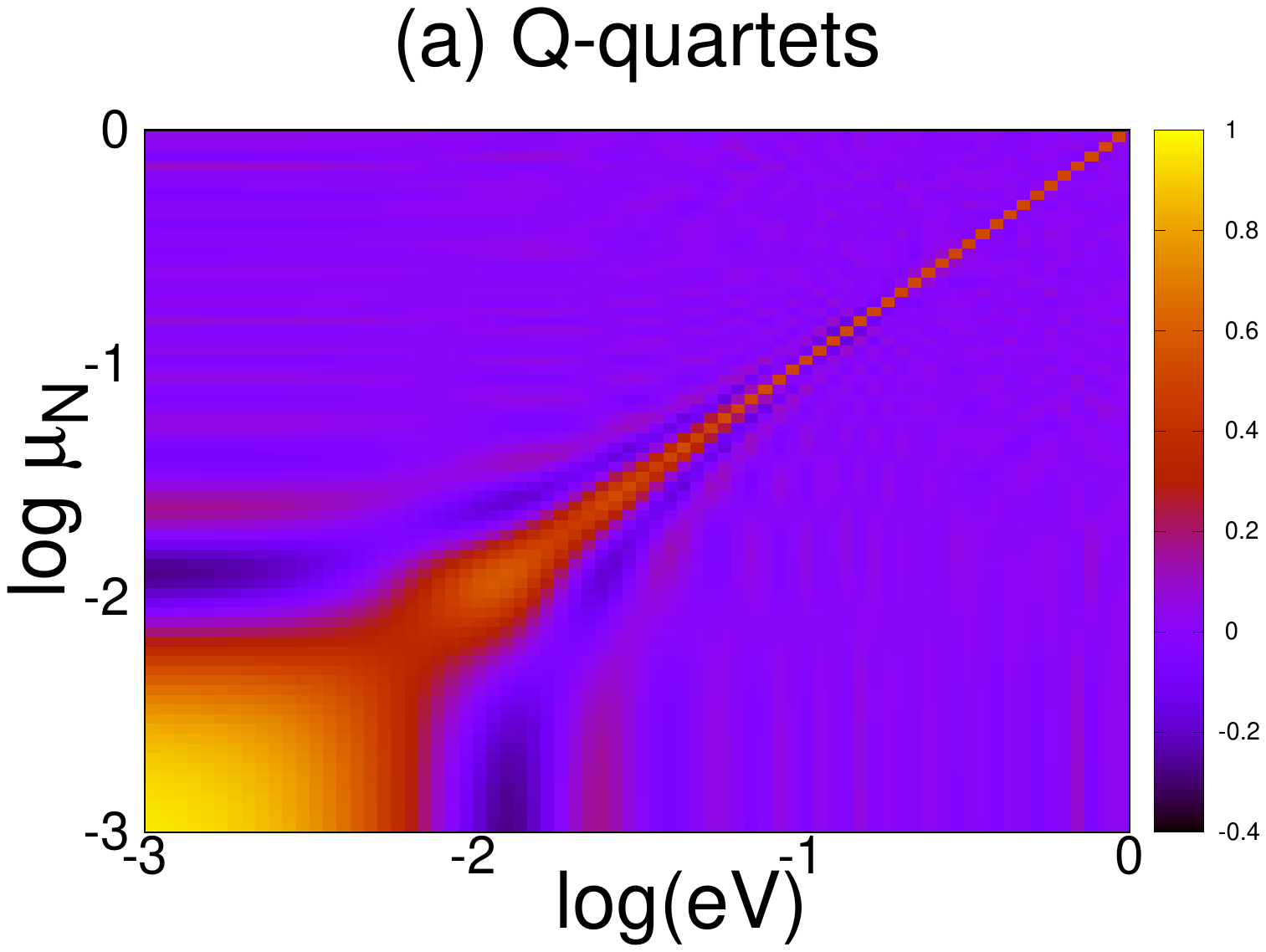} \includegraphics[width=.3\textwidth]{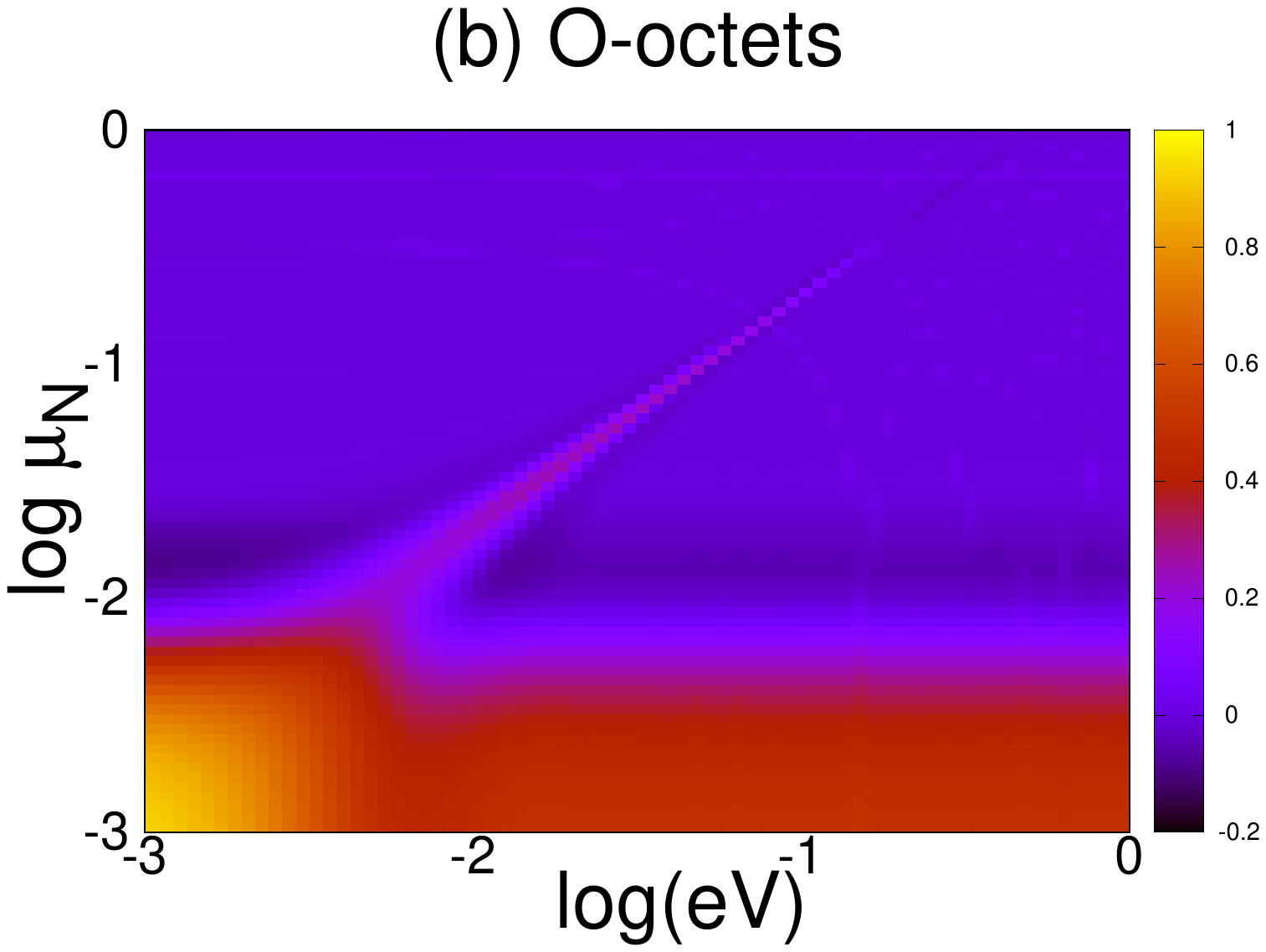} \includegraphics[width=.3\textwidth]{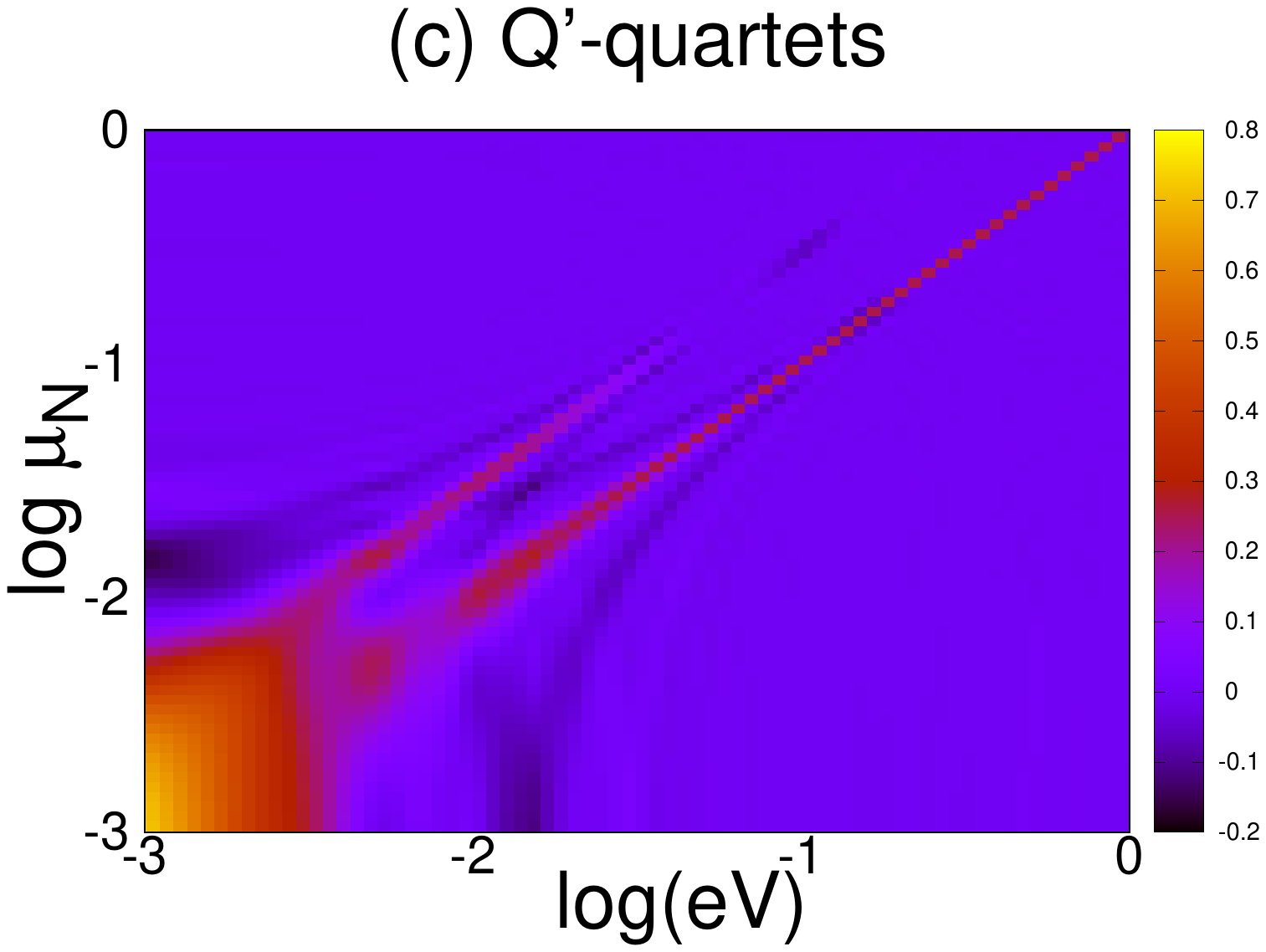}}

  \centerline{\includegraphics[width=.3\textwidth]{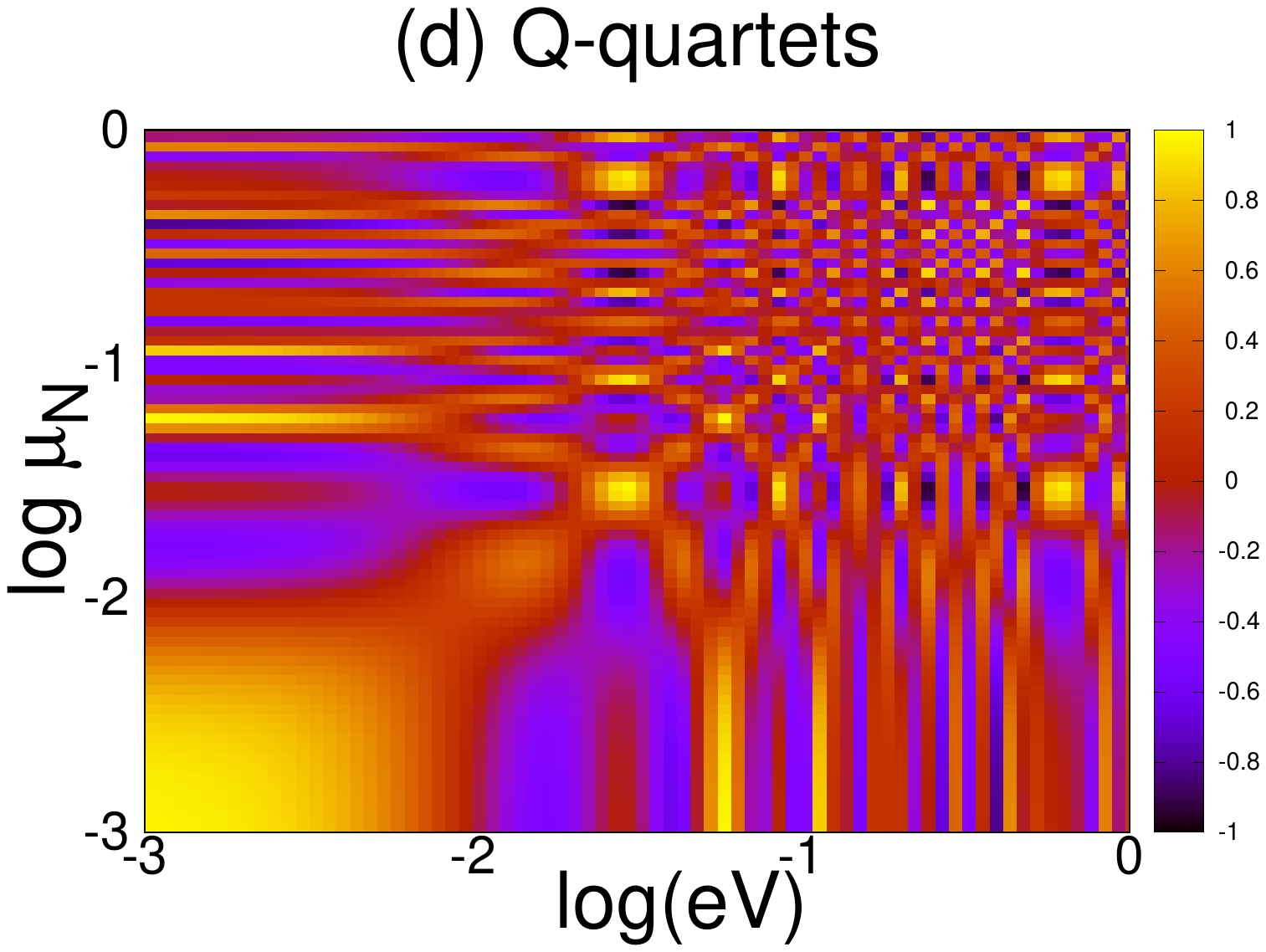} \includegraphics[width=.3\textwidth]{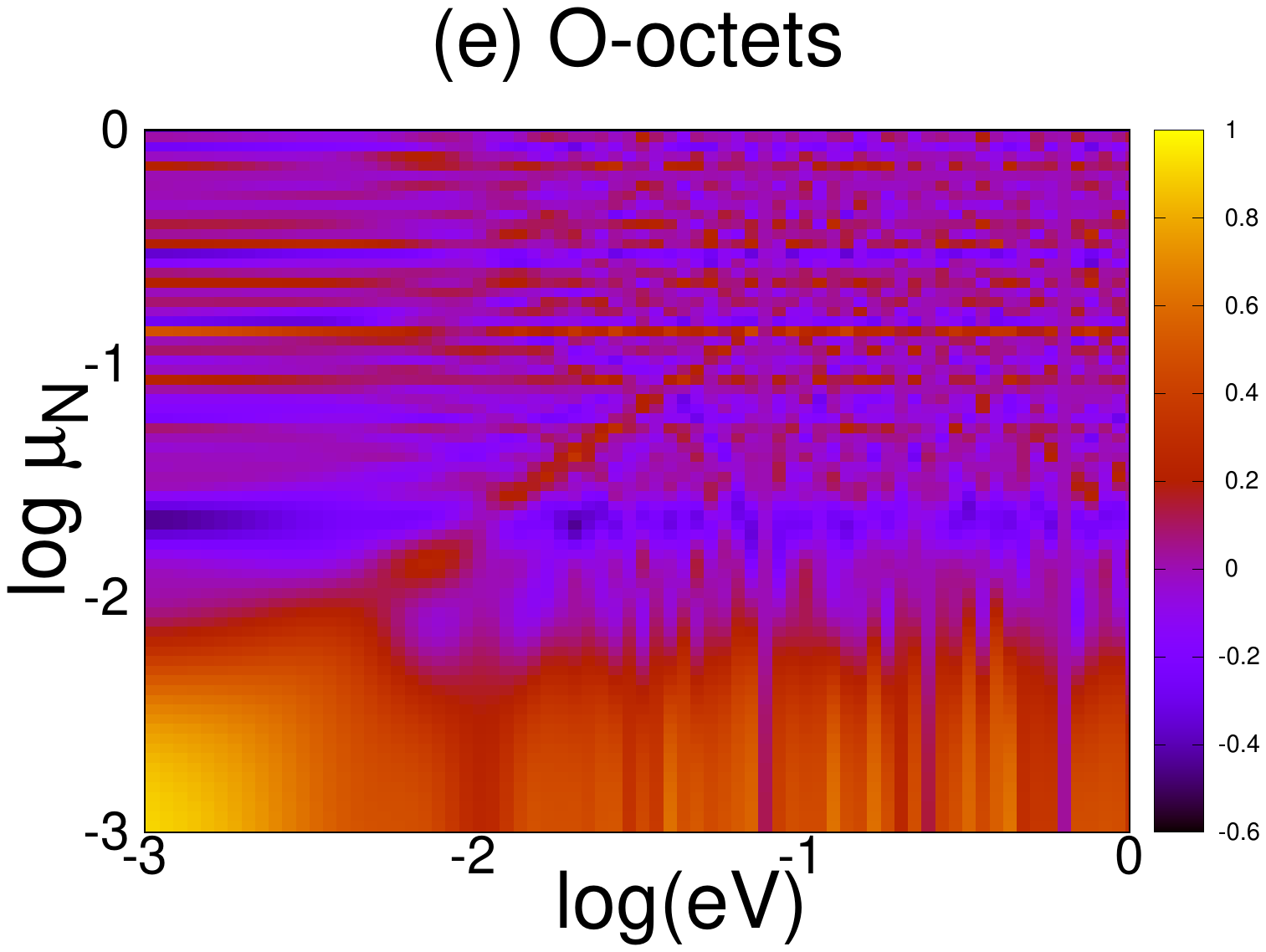} \includegraphics[width=.3\textwidth]{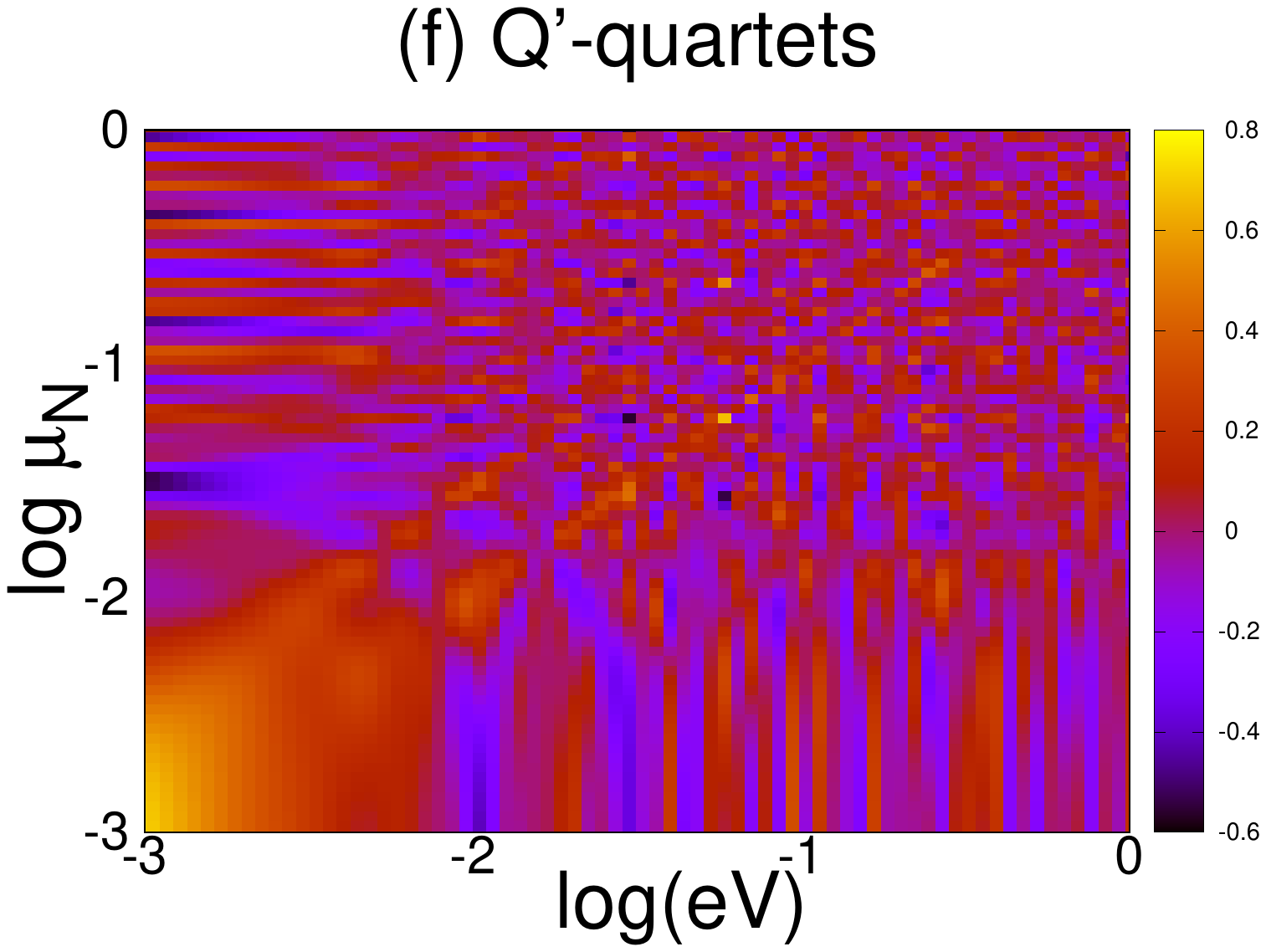}}
\end{minipage}
  \caption{{\it The Floquet-MAR {finite-bias}
      {resonances}} visible in the dimensionless
    current susceptibility $\partial i'(eV,\mu_N).\partial \mu_N$, as
    a function of $(\log (eV),\,\log \mu_N)$ on the $(x,y)$ axis, for
    the $Q$-quartets (a) and (d), the $O$-octets (b) and (e) and the
    $Q'$-quartets (c) and (f). The summation of the microscopic $Q$,
    $O$ and $Q'$ diagrams over the interfaces, is carried out on
    panels (a), (b) and (c) taking into account nonlocal propagation
    in the 2D conductor continuum. The contacts form the $L\times L$
    square-shape geometry in Fig.~\ref{fig:1}a, with $L=10$. Panel
    (d), (e) and (f) corresponds to maps with averaging over the
    $\lambda_F$-oscillations in single channels. The colorscale axis is
    in arbitrary units.
  \label{fig:multiplot2}}
\end{figure*}

{\it Elementary quartets as Floquet-MAR dimers:} The standard
lowest-order ${Q}$-quartet diagram is shown in Fig.~\ref{fig:2}c and
Figs.~\ref{fig:diagrammes}a-c. Introducing the spectral current at the
energy $\omega$, the elementary quartet diagram in
Fig.~\ref{fig:diagrammes}b produces spectral weight at the energies
$\Omega_L^{(\pm)}=\pm eV+\omega$ and $\Omega_R^{(\mp)}=\mp eV-\omega$,
thus with four Floquet-MAR {resonances in the spectral} {current
  susceptibility} at finite $\omega$. The dimensionless spectral
current takes the $i_{{Q}}(\omega) \cos\varphi_q$ form, where the
$\cos \varphi_q$ sensitivity relates to the so-called $\cos \Phi$
Aharonov-Bohm component of the current in Andreev interferometers,
where $\Phi$ is the flux enclosed by the loop
\cite{Melin2024,Zaikin1,Zaikin2,Zaikin3,Zaikin4}. The dimensionless
{equilibrium spectral current $i_Q(\omega)$ is
  evaluated as the current entering $S_L$, and results from double
  Cooper pair splitting from $S_B$ to $(S_L,\,S_R)$. The dimensionless
  equilibrium spectral current kernel $i_Q(\omega)$ through the
  $S_L$-$N$ interface is deduced from the equilibrium Keldysh Green's
  function calculations with $\mu_N=eV=0$:}
\begin{equation}
  \label{eq:I-calQ}
  i_Q(\omega)=\prod_{\langle k,l \rangle} \frac{1}{k_F R_{k,l}}
  \cos\left(\frac{2\omega R_{k,l}}{\hbar v_F}\right) ,
\end{equation}
with the corresponding transmission mode labeling in
Fig.~\ref{fig:diagrammes}a, i.e. $\langle k,\,l \rangle=\langle
\alpha,\,\gamma\rangle,\, \langle \gamma,\,\beta\rangle$. {Throughout,
  $\varphi_Q=\varphi_L+\varphi_R-2\varphi_B$. The equilibrium physical
  current is obtained by integrating the $i_Q(\omega)$ kernel with the
  occupation function and multiplying it by the tunneling prefactor
  and $\cos \varphi_Q $, see section IV of the SM \cite{supinfo}. At
  finite $V$ and $\mu_N$, $i _ Q(eV,\mu_N)$ denotes the corresponding
  dimensionless energy-integrated nonequilibrium current, while
  $\partial i'_Q /\partial \mu_N$ is its current susceptibility. The
  measured transconductance with respect to $V_P$ is $G_{L,P} = dI_L/d
  V_P=(\partial I_L/\partial \mu_N)\times(\partial \mu_N/\partial
  V_P)+ \partial I_L/\partial V_P$. The notation $R_{k,l}$ in
  Eq.~(\ref{eq:I-calQ})} stands for the separation between the
contacts, see Fig.~\ref{fig:1}b. {As a simplified model, we first
  treat the inter-contact distances $R_{k,l}$ as independent random
  variables and average over their distribution}
\begin{equation}
  \label{eq:I-calQav}
  \overline{i(\omega)}=\left(\int {\cal
    P}\left(R_{k,l}\right) \frac{1}{k_F R_{k,l}}
  \cos\left(\frac{2\omega R_{k,l}}{\hbar v_F}\right) d
  R_{k,l}\right)^N
  .
\end{equation}
The flat ${\cal P}(R_{k,l})$ simulates the strong proximity effect,
and leads to the {finite-bias resonances in the current
  susceptibility} at the energies $\Omega_L^{(\pm)}=\pm eV$ and
$\Omega_R^{(\mp)}=\mp eV$, see Fig.~\ref{fig:2}b. Considering now the
more realistic square-shape ballistic $L\times L$ 2D conductor (with
$L=10$ in Fig.~\ref{fig:multiplot2}), the zero-energy {resonance} in
the spectral current $i_Q(\omega)$ \cite{Beenakker,Floser} fixes the
width $|\omega|\alt\omega_c$ {of the Floquet-MAR finite-bias current
  susceptibility} {resonance} in the {nonequilibrium}
spectral current $i'_Q(eV,\mu_N) \cos\varphi_q$ {at
  finite $eV$ and $\mu_N$,} with $\omega_c$ the analogous Thouless
energy, see Fig.~\ref{fig:diagrammes}c.

The dimensionless quartet current susceptibility $\chi_I$ quantifies
how the quartet current responds to changes in the electrochemical
potential. The perturbative calculations, see the SM \cite{supinfo},
nontrivially relate the nonequilibrium $\chi_I$ to the Floquet replica
of the equilibrium spectral current given by
Eq.~(\ref{eq:I-calQ}). Specifically, the Keldysh Green's functions
yield the following form of the dimensionless current susceptibility,
see section V of the SM \cite{supinfo}:
\begin{eqnarray}
  \label{eq:conductance1}
  \frac{\partial i'_Q(eV,\mu_N)}{\partial \mu_N} =
  \sum_{\tau=\pm} i_{{Q}}\left(2eV+2\tau\mu_N\right)
  ,
\end{eqnarray}
{see Eq.~(45) in the SM \cite{supinfo}.} The emerging
          {finite-bias Floquet-MAR resonances in the current
            susceptibility} at $\mu_N=\pm eV$, see Fig.~\ref{fig:2}b,
          are compatible with Fig.~\ref{fig:multiplot2}a calculated
          for $eV>0$ and $\mu_N>0$.  Now, we address the higher-order
          Floquet-MAR multiplets.

{\it Higher-order Floquet-MAR multiplets:} Neglecting the
Landau-Zener-Stückelberg interference in the large-scale ballistic 2D
conductor-based MJJs, the Floquet-MARs are viewed as dressing the quartet
diagrams at higher-order in tunneling. Tunneling from $S_L$ into the
ballistic 2D conductor $N$ is dressed by noting that the insertion of a
{\it MAR vertex correction} across $S_B$ changes the energy from
$\omega$ to $\omega\pm2eV$. An inverse-MAR vertex correction inserted
at the opposite interface changes the energy from $\omega\pm2eV$ back
to $\omega$, see Fig.~\ref{fig:2}d and Figs.~\ref{fig:diagrammes}d, e.
The resulting overall energy-conserving higher-order snake diagrams
form the ${O}$-octets and ${Q}'$-quartets, and they yield the
DC-current susceptibility of the corresponding Floquet-MARs, see
Fig.~\ref{fig:2}d and Fig.~\ref{fig:2}e. The dimensionless spectral
currents $i_{{O}}(\omega)\cos(2\varphi_q)$ and
$i_{{Q}'}(\omega)\cos\varphi_q$ are still given by
Eq.~(\ref{eq:I-calQ}), but now with the four $\langle k,l\rangle$
associated to the corresponding snake diagrams.

Our physical framework is now extended to the higher-order multipair
processes that emerge from the higher-order snake
diagrams{. The quartets and octets appear at the
  orders $8$ and $16$ in perturbation theory in tunneling, thus with
  the typical ratio $\sim T_N^4/(k_F R_0)^2$ of their corresponding
  anomalies, where $T_N$ is the normal-state transparency and $R_0$
  the typical distance between the contacts. However, increasing the
  contact transparencies beyond perturbation theory for $T_N\alt 1$ is
  likely to enhance the signal of the higher-order resonances, while
  preserving their location in the $(eV,\,\mu_N)$ plane.}  We
demonstrate that the corresponding $O$-octets and $Q'$-quartets
produce the more complex spectra in Fig.~\ref{fig:2}d, f, and we
calculate the corresponding distinguishing features in the maps of the
DC-current susceptibility, see Figs.~\ref{fig:multiplot2}b-c,
e-f. Similarly to Eq.~(\ref{eq:conductance1}), the calculations
presented in sections VI and VII of the SM lead to express the
$O$-octet and $Q'$-quartet dimensionless current susceptibilities
$\left(\partial i'_{{O}}/\partial \mu_N\right)\cos(2\varphi_q)$ and
$\left(\partial i'_{{Q}'}/\partial \mu_N\right)\cos\varphi_q$ as a
summation over the Floquet replica of the corresponding spectral
currents, {see Eqs.~(60) and (70) in the SM
  \cite{supinfo}:}
\begin{eqnarray}
  \label{eq:conductance2}
   \frac{\partial i'_{{O}}(eV,\mu_N)}{\partial \mu_N} &&=
  2 i_{{O}}(2\mu_N)+
  \sum_{\tau=\pm} i_{O}(4eV+2\tau\mu_N)\\ \label{eq:conductance3}
  \frac{\partial
    i'_{{Q}'}(eV,\mu_N)}{\partial \mu_N}&& =\\
&&    \sum_{\tau=\pm} \left[ i_{{Q}'}(2eV+2\tau\mu_N)
    + i_{{Q}'}(6eV+2\tau\mu_N)\right]
  \nonumber
  ,
\end{eqnarray}
with the Floquet-MAR trimer $\mu_N=0,\,\pm 2eV$ for the ${O}$-octets
(see Fig.~\ref{fig:2}d) and the Floquet-MAR quadruplet $\mu_N=\pm
eV,\,\pm 3eV$ for the ${Q}'$-quartets (see Figs.~\ref{fig:2}f). The
numerical calculations in Figs.~\ref{fig:multiplot2}b and
Fig.~\ref{fig:multiplot2}c are restricted to the $eV>0,\,\mu_N>0$
quadrant and they reveal the expected $\mu_N=0,\,2eV$ Floquet-MAR
{finite-bias} {resonances in the current susceptibility, see}
Fig.~\ref{fig:multiplot2}b, and $\mu_N=eV,\,3eV$ in
Fig.~\ref{fig:multiplot2}c. {Thus, the main features of the current
  susceptibility maps in Figs.~\ref{fig:multiplot2}a-c can be captured
  to a large extent by the {\it rule of the thumb} of the energy
  diagrams in Figs.~\ref{fig:2}b, d, f.} {Due to
  energy conservation, the octets resonance line is superimposed with
  the quartets in the $(V_L,\,V_R)$ plane of the bias
  voltages. However, the degeneracy is lifted in the plane of the
  electrochemical potential and bias voltage $V\equiv V_L=-V_R$,
  making possible the observation of the octets.}
  
{\it Localized single-channel contacts:} {One-dimensional (1D)
  configurations turned out to be useful to demonstrate the zero-bias
  conductance peaks in two- and multiterminal superconducting hybrids
  \cite{Beenakker,Floser}. There, disorder is simply introduced by
  averaging the current or noise over the separation $R_{k,l}$ between
  the contacts, within the
  $\left[R_{0,k,l}-\lambda_F/2,R_{0,k,l}+\lambda_F/2\right]$
  window. In order to bridge with these 1D approaches, we now take
  each of the $S_L$-$N$, $S_R$-$N$ and $S_B$-$N$ contact to be
  localized on a single tight-binding tunneling amplitude, and
  calculate the corresponding maps of the current susceptibility, see
  Figs.~\ref{fig:multiplot2}d-f, still averaging over the short-scale
  $\lambda_F$-oscillations \cite{Beenakker,Floser}.  The resulting
  Floquet-MAR finite-bias current susceptibility resonance spectra are
  shown on panels d-f. Those single-channel resonances are also
  captured by a simplified toy-model, see section IX of the SM
  \cite{supinfo}.}

{\it Quantum noise:} {The dissipative components of the
  current, proportional to the cosine of the gauge-invariant
  superconducting phase combination, are central to the device on
  Fig.~\ref{fig:1}b defined on an infinite 2D conductor. We now argue
  that the quantum noise cross-correlations are proportional to this
  dissipative current. This will demonstrate the granularity and
  quantum coherence of the Floquet-MAR resonances at finite bias, via
  an evaluation of the corresponding finite Fano factor. Specifically,
  we calculate} the Fourier transform of the current-current
cross-correlation expectation value $S_{L,R}(t,t')=\langle\delta
\hat{I}_L(t) \delta \hat{I}_R(t')\rangle$ between the
$\delta\hat{I}_L(t)=\hat{I}_L(t)-\langle \hat{I}_L \rangle$ and the
$\delta\hat{I}_R(t')= \hat{I}_R(t')-\langle \hat{I}_R\rangle$ current
fluctuations at the left and right contacts. We use the Keldysh
Green's functions to evaluate the average currents $I_L$ and $I_R$,
and the current-current cross-correlations $S_{L,R}$, starting with a
simple ballistic Andreev interferometer \cite{Melin2024}. We also
generated the expressions of the {the noise for the
  Andreev tubes associated to} the snake diagrams of order $N$, which
yields the following Fano factor, see section VIII in the SM
\cite{supinfo}:
\begin{equation}
  \label{eq:F}
  F_N\equiv\frac{\partial S_{L,R}(eV,\mu_N)/\partial
      \mu_N}{\partial I_L(eV,\mu_N)/\partial \mu_N}=2N.
\end{equation}
To demonstrate Eq.~(\ref{eq:F}), we express the noise kernel and the
current in terms of the fully dressed $\hat{G}^{+,-}$ and
$\hat{G}^{-,+}$ Nambu-Keldysh Green's functions, see section IIC in
the SM \cite{supinfo}, and expand in perturbation in the tunneling
amplitudes at the lowest order. The noise and the current
susceptibilities are then expressed as $\partial
I_L(eV,\mu_N)/\partial \mu_N\simeq A_N X$ and $\partial
S_{L,R}(eV,\mu_N)/\partial \mu_N\simeq B_N X$, where the common factor
$X$ is proportional to the product of the pairs of the nonlocal
Green's functions connecting $S_L$ to $S_R$ within each snake diagram
of order-$N$. Section VIIIA in the SM \cite{supinfo} establishes that
$B_N/A_N=2N$. {In summary, Eq.~(\ref{eq:F}) demonstrates that the
  quantum noise cross-correlations are proportional to the dissipative
  current, that, in turn is proportional to the voltage $V$
  because $\mu_N \propto V$ if the populations can freely adjust their
  electrochemical potential, and the coupling to the leads is
  generally nonsymmetrical. The resulting {linear-in-$V$} quantum
  noise is orders of magnitude larger than the exponentially
  small-in-$V$} Landau-Zener-Stückelberg quantum noise
cross-correlations of the superconducting quantum dots
\cite{Melin-Sotto}.

{\it Conclusions:} {We have developed a microscopic
  Floquet-MAR description for voltage-biased multiterminal Josephson
  junctions formed on an extended ballistic two-dimensional
  conductor. Unlike earlier treatments based primarily on discrete
  Andreev levels in quantum dots or finite systems, the present theory
  demonstrates that a continuous electronic spectrum supports
  resolvable Floquet-MAR structures. The independent
  electrochemical-potential coordinate $\mu_N$ produces distinct
  susceptibility resonances for elementary quartets, octets, and
  higher-order quartets and removes the quartet–octet degeneracy
  present in conventional bias-voltage maps. The accompanying noise
  response provides an additional diagnostic of correlated multipair
  transport. The interplay between the Floquet-MAR periodic orbits and
  the quartet winding number resonances \cite{Rashid2026} is a
  promising perspective, as well as the interplay with the topology in
  the sense of the Berry curvature or Chern numbers
  \cite{Riwar2016,Eriksson2017,Gavensky2018,Repin2019,Fatemi2021,Melo2022}.}
 
{\it Acknowledgements:} R.M. wishes to express his gratitude to
Beno\^{\i}t Dou\c{c}ot, Katie Huang, Philip Kim and Yuval Ronen for
their previous collaboration on the Floquet theory of MJJs. R.M. and
R.D. acknowledge the financial support from the SUPRADEVMAT
International Research Project between the French CNRS-Grenoble and
the German KIT-Karlsruhe. R.D. acknowledges the funding from the
Deutsche Forschungsgemeinschaft (DFG, German Research Foundation) --
467596333 and the support from the Helmholtz Association through
program NACIP. M.K. acknowledges funding from the Pennsylvania State
University Materials Research Science and Engineering Center supported
by the US National Science Foundation (DMR 2011839) and the US
National Science Foundation (DMR 2415756).

\end{document}


\title{Floquet-Multiple Andreev Reflections:\\Supplemental Material}

\author{R\'egis M\'elin}
\email{regis.melin@neel.cnrs.fr}

\affiliation{Universit\'e Grenoble-Alpes, CNRS, Grenoble INP, Institut
 NEEL, Grenoble, France}

\author{Romain Danneau}

\affiliation{Institute for Quantum Materials and Technologies,
  Karlsruhe Institute of Technology, Karlsruhe D-76021, Germany}

\author{Morteza Kayyalha}

\affiliation{Department of Electrical Engineering, The Pennsylvania
State University, University Park, Pennsylvania 16802, USA}

\begin{abstract}
  The Supplemental Material summarizes the technical details of the
  calculations.
\end{abstract}

\maketitle

The Supplemental Material (SM) is organized as follows. The
Hamiltonians are presented in section~\ref{sec:H}. The methods are
presented in section~\ref{sec:methods}. General symmetry arguments for
the current are presented in section~\ref{sec:general}. The simplest
two-terminal DC-Josephson effect is discussed in
section~\ref{sec:2T-eq}, which allows demonstrating Eq.~(1) in the
main text. The $Q$-quartets, $O$-octets and $Q'$-quartets are next
treated in sections~\ref{sec:Q},~\ref{sec:O} and~\ref{sec:Q'}
respectively, which demonstrates Eqs.~(3)-(5) in the main text. The
Fano factor is calculated in section~\ref{sec:Fano}, thus providing a
demonstration of Eq.~(6) in the main text. Finally, a toy-model for
the emerging Floquet resonances in the current susceptibility is
presented in section~\ref{sec:toy} of the SM, which supports Fig. 4d-f
in the main text.

\section{Hamiltonians}
\label{sec:H}
In this section of the SM, we provide the BCS Hamiltonian for the
superconductors (see subsection~\ref{sec:H-BCS}), the 2D conductor
Hamiltonian (see subsection~\ref{sec:2D}) and the hopping amplitude
between both (see subsection~\ref{sec:hopping}).

\subsection{BCS Hamiltonian}
\label{sec:H-BCS}
In this subsection of the SM, we introduce the BCS Hamiltonian:
\begin{eqnarray}
 \label{eq:H-BCS}
 {\cal H}_{BCS} = - W \sum_{\langle x,y\rangle} \sum_\sigma
 \left( c_{x,\sigma}^+ c_{y,\sigma} + c_{y,\sigma}^+ c_{x,\sigma}\right)
 -\sum_x \left( \Delta_x
 \exp(i\varphi_x) c_{x,\uparrow}^+ c_{x,\downarrow}^+ +
 \Delta_x \exp(-i\varphi_x) c_{x,\downarrow} c_{x,\uparrow}
 \right)
 ,
\end{eqnarray}
where $x,y$ and $x$ in both terms of Eq.~(\ref{eq:H-BCS}) run over the
tight-binding sites $x$ and $y$, and $\langle x,y\rangle$ stands for
neighboring tight-binding sites. The notation $W$ stands for the
band-width and $(\Delta_x,\varphi_x)$ are the local superconducting
amplitudes and phases at the tight-binding site~$x$.

\subsection{2D conductor Hamiltonian}
\label{sec:2D}

In this subsection of the SM, we introduce the 2D conductor Hamiltonian:
\begin{equation}
 \label{eq:H-tb-2D}
 {\cal H}_{2D,N}=-W \sum_{\langle x,y \rangle} \sum_\sigma
 \left(c_{x,\sigma}^+ c_{y,\sigma} + c_{y,\sigma}^+
 c_{x,\sigma}\right) ,
\end{equation}
where $x$ and $y$ run over the corresponding tight-binding lattice
sites, as in the above Eq.~(\ref{eq:H-BCS}).

\subsection{Interfacial hopping Hamiltonian}
\label{sec:hopping}

In this subsection of the SM, we introduce the coupling between the
superconductors and the normal conductors:
\begin{equation}
 {\cal H}_{tun}= - \Sigma_0 \sum_{x,y} \sum_\sigma \left(
 c_{x,\sigma}^+ c_{y,\sigma} + c_{y,\sigma}^+
 c_{x,\sigma}\right) ,
\end{equation}
where, again, $x$ and $y$ run over the interfacial tight-binding
sites.

\section{Methods}
\label{sec:methods}
In this section of the SM, we introduce how the current and the noise
are calculated using the Keldysh Green's functions, see the pioneering
Refs.~\onlinecite{Caroli1,Caroli2,Cuevas,Cuevas-noise,Averin,Averin-noise}. Subsection~\ref{subsecA}
presents the expression of the nonequilibrium
current. Subsection~\ref{subsecB} specializes to a device that is
grounded but phase-biased. Subsection~\ref{subsecC} present the
general formula for the quantum noise
cross-correlations. Subsection~\ref{subsecD} deals with how the
Andreev pair propagation averages over the oscillations at the
small-scale of the Fermi wave-length.

\subsection{Expression for the nonequilibrium current}
\label{subsecA}

In this subsection of the SM, we explain how the current is calculated from the
Keldysh Green's function. The superconducting Green's functions have
two Nambu components: ``1'' for spin-up electron and ``2'' for
spin-down hole, that are equivalently denoted by ``e'' and ``h'' in
the main text. We consider that the nonsuperconducting central region
$N$ is connected by tight-binding amplitudes $\hat{\Sigma}$ to an
arbitrary number of superconducting leads, generally denoted by
$S$. Then, the fully dressed advanced and retarded Green's functions
$\hat{G}^A$ and $\hat{G}^R$ are the solutions of the following Dyson
equations:
\begin{equation}
  \label{eq:D}
\hat{G}^{A,R} = \hat{g}^{A,R} + \hat{g}^{A,R} \hat{\Sigma}
\hat{G}^{A,R}.
\end{equation}
The Keldysh
Green's function $\hat{G}^{+,-}$ is obtained from the Dyson-Keldysh
equation:
\begin{equation}
  \label{eq:Gpm}
  \hat{G}^{+,-}=\left(\hat{I}+\hat{G}^R \hat{\Sigma}\right)
  \hat{g}^{+,-}
  \left(\hat{I} + \hat{\Sigma} \hat{G}^A\right)
  .
\end{equation}
We denote by $\Sigma_{a,\alpha}=\Sigma_{\alpha,a}$ the tight-binding
amplitude forming electron and hole transmission between the $a$ and
$\alpha$ tight-binding sites at the interfaces of the left
superconductor $S_L$ and the right normal metal $N$, respectively, see
Figure 3a in the main text for the notations $a$ and $\alpha$. The
spectral current flowing between $S_L$ and $N$ is related to the
DC-component of the Keldysh Green's function:
\begin{eqnarray}
  \label{eq:A1}
  I(\omega)&=&\mbox{Nambu-trace}\left\{\hat{\tau_z}\left[\Sigma_{a,\alpha}
    \hat{G}^{+,-}_{\alpha,a} - \hat{\Sigma}_{\alpha,a}
    \hat{G}^{+,-}_{a,\alpha}\right]\right\}_{DC}\\ \nonumber&=& \left\{
  \hat{\Sigma}_{a,\alpha}^{1,1} \hat{G}^{+,-,1,1}_{\alpha,a}
  - \hat{\Sigma}_{a,\alpha}^{2,2} \hat{G}^{+,-,2,2}_{\alpha,a}
  -\hat{\Sigma}_{\alpha,a}^{1,1} \hat{G}^{+,-,1,1}_{a,\alpha}
  +\hat{\Sigma}_{\alpha,a}^{2,2}
  \hat{G}^{+,-,2,2}_{a,\alpha}\right\}_{DC} ,
\end{eqnarray}
where the Pauli matrix $\hat{\tau}_z$ acts in the Nambu
space. Eq.~(\ref{eq:D}) can be iterated according to
\begin{eqnarray}
  \hat{G}&=& \hat{g}+\hat{g}\hat{\Sigma}\hat{G} =
  \hat{g}+\hat{g}\hat{\Sigma}\hat{g}+\hat{g}\hat{\Sigma}
  \hat{g}\hat{\Sigma}\hat{G} =
  \hat{g}+\hat{g}\hat{\Sigma}\hat{g}+\hat{g}\hat{\Sigma}
  \hat{g}\hat{\Sigma}\hat{g} +
  \hat{g}\hat{\Sigma}\hat{g}\hat{\Sigma}\hat{g}\hat{\Sigma} \hat{G} =
  ...
\end{eqnarray}
Truncating the series at a finite order produces perturbation theory
of the Green's function $\hat{G}$ in powers of $\Sigma^2$. Similarly
proceeding with the Keldysh Green's function given by
Eq.~(\ref{eq:Gpm}) produces an expansion of the current given by
Eq.~(\ref{eq:A1}), that can be visualized as diagrams where the
nonlocal Green's functions across $N$ are gathered in a pairwise
manner, in such a way as to select the semiclassical paths that
constructively interfere in the multichannel summation, i.e. the
averaging of the current is over the oscillations at the short scale
of the Fermi wave-length $\lambda_F$.

\subsection{Expression for the equilibrium supercurrent}
\label{subsecB}

In this subsection of the SM, we specialize Eq.~(\ref{eq:A1}) to
devices that are biased in phase and grounded. At equilibrium, the
spectral supercurrent ${\cal I}(\omega)$ is deduced from the above
Eq.~(\ref{eq:A1}):
\begin{eqnarray}
  \label{eq:I-omega-1}
  {\cal I}(\omega)&=&\mbox{Nambu-trace}\left[\hat{\sigma}_z
    \left(\hat{\Sigma}_{a,\alpha} \hat{G}^{+,-}_{\alpha,a,eq} -
    \hat{\Sigma}_{\alpha,a}
    \hat{G}^{+,-}_{a,\alpha,eq}\right)\right]\\ &=& n_F(\omega)
  \mbox{Nambu-trace}\left[\hat{\sigma}_z \left[\hat{\Sigma}_{a,\alpha}
      \left(G^A_{\alpha,a}-\hat{G}^R_{\alpha,a})\right] -
    \hat{\Sigma}_{\alpha,a} \left(
    \hat{G}^A_{a,\alpha}-\hat{G}^R_{a,\alpha}\right)\right]\right]
\label{eq:I-omega-2}
,
\end{eqnarray}
where we used the following form of the equilibrium Keldysh Green's function:
\begin{equation}
  \label{eq:Gpm-eq}
  \hat{G}^{+,-}_{eq}=
  n_F(\omega)\left(\hat{G}^A(\omega)-\hat{G}^R(\omega)\right)
  .
\end{equation}
In this equation, the equilibrium Fermi-Dirac distribution function
$n_F(\omega)$ is such that $n_F(\omega)=\theta(-\omega)$ in the
considered limit of zero temperature, where the notation $\theta(x)$
is used for the Heaviside function, taking the value $\theta(x)=1$ if
$x<0$ and $\theta(x)=0$ if $x>0$.

\subsection{Expression for the quantum noise cross-correlations}
\label{subsecC}

In this subsection of the SM, we provide the expression of the kernel
$\hat{K}_{a,b}$ of the quantum noise cross-correlations
$S_{a,b}=e^2/\hbar^3 \mbox{Tr}\left[K_{a,b}\right]$:
\begin{eqnarray}
  \label{eq:K}
  \hat{K}_{a,b}&=& \hat{\Sigma}_{\beta,b} \hat{\tau}_3
  \hat{G}^{+,-}_{b,a} \hat{\Sigma}_{a,\alpha} \hat{\tau}_3 \hat{G}^{-,+}_{\alpha,\beta}
  =
  \hat{\Sigma}_{b,\beta} \hat{\tau}_3 \hat{G}^{+,-}_{\beta,\alpha} \hat{\Sigma}_{\alpha,a}
  \hat{\tau}_3 \hat{G}^{-,+}_{a,b}
  - \hat{\Sigma}_{\beta,b} \hat{\tau}_3 \hat{G}^{+,-}_{b,\alpha} \hat{\Sigma}_{\alpha,a}
  \hat{\tau}_3 \hat{G}^{-,+}_{a,\beta}
  - \hat{\Sigma}_{b,\beta} \hat{\tau}_3 \hat{G}^{+,-}_{\beta,a} \hat{\Sigma}_{a,\alpha}
  \hat{\tau}_3 \hat{G}^{-,+}_{\alpha,b}
  + \left(\tau \leftrightarrow \tau' \right)
  .
\end{eqnarray}
This expression is used to evaluate the ratio between the current
noise cross-correlations and the current, which is known as the Fano
factor, see also section~\ref{sec:Fano} of the SM.

\subsection{Andreev pair transmission across a ballistic 2D conductor}
\label{subsecD}

In this section of the SM, we calculate how the transmission modes
across a ballistic 2D normal conductor average out over the
short-scale Fermi oscillations. The nonlocal advanced electron and
hole Nambu Green's functions of a ballistic 2D conductor at the
energies $\omega_1$ and $\omega_2$ are given by
\begin{eqnarray}
  \label{eq:J0-1}
  \hat{g}_{\alpha,\beta}^{A,1,1}&=&\frac{i}{W}
  J_0\left[\left(k_F+\frac{\omega_1}{\hbar v_F}\right)
    R_{\alpha,\beta}\right]\\
  \hat{g}_{\alpha,\beta}^{A,2,2}&=&\frac{i}{W}
  J_0\left[\left(k_F-\frac{\omega_2}{\hbar v_F}\right)
    R_{\alpha,\beta}\right]
  ,
  \label{eq:J0-2}
\end{eqnarray}
where $W$ is the band-width, $v_F$ the Fermi velocity and
$R_{\alpha,\beta}$ is the separation between the tight-binding sites
$\alpha$ and $\beta$, belonging to the 2D conductor. The notation
$J_0$ in Eqs.~(\ref{eq:J0-1})-(\ref{eq:J0-2}) stands for one of the
Bessel functions. Eqs.~(\ref{eq:J0-1})-(\ref{eq:J0-2}) have the
following expression at large $k_F R_{\alpha,\beta}\gg 1$:
\begin{eqnarray}
  \label{eq:g2D-1}
  \hat{g}_{\alpha,\beta}^{A,1,1/2,2}\left(R_{\alpha,\beta},\omega\right)
  &\simeq&\frac{i}{W\sqrt{k_F R_{\alpha,\beta}}}
  f^{(\pm)}_{\alpha,\beta}(\omega)
  ,
\end{eqnarray}
where we neglected the energy-dependence of the Fermi wave-vector in
the slowly-varying geometrical prefactor and
\begin{equation}
  \label{eq:fpm}
  f^{(\pm)}_{\alpha,\beta}(\omega)=
    \cos \left[\left(k_F\pm\frac{\omega}{\hbar v_F}\right) R_{\alpha,\beta}-\frac{\pi}{4}\right]
    .
\end{equation}

In the limit of small separation $R_{\alpha,\beta}\rightarrow 0$, the advanced Green's functions
given by Eqs.~(\ref{eq:J0-1})-(\ref{eq:J0-2}) go to the following standard expression
of the local Green's function:
\begin{equation}
  \label{eq:rho-0}
  \hat{g}_{loc}^A=i\pi \rho^{(0)}_{loc}, \mbox{where }
  \rho^{(0)}_{loc}=\frac{1}{W} .
\end{equation}

Evaluating the products of the $1,1$ and $2,2$ components of
$\hat{g}_{\alpha,\beta}^{A,1,1/2,2}\left(R_{\alpha,\beta},\omega\right)$,
see Eqs.~(\ref{eq:g2D-1})-(\ref{eq:fpm}), leads to
\begin{eqnarray}
  \label{eq:g-g-ball}
  &&
  \hat{g}_{\alpha,\beta}^{A,1,1}\left(R_{\alpha,\beta},\omega_1\right)
  \hat{g}_{\beta,\alpha}^{A,2,2}\left(R_{\alpha,\beta},\omega_2\right)=\\
  \nonumber
  &&
  -\frac{1}{W^2(k_F R_{\alpha,\beta})}
  \left[\cos\left(k_F R_{\alpha,\beta}-\frac{\pi}{4}\right)
    \cos\left(\frac{\omega_1 R_{\alpha,\beta}}{\hbar v_F}\right)
    -
    \sin\left(k_F R_{\alpha,\beta}-\frac{\pi}{4}\right)
    \sin\left(\frac{\omega_1 R_{\alpha,\beta}}{\hbar v_F}\right)
    \right]\times\\\nonumber
  &&
  \left[\cos\left(k_F R_{\alpha,\beta}-\frac{\pi}{4}\right)
    \cos\left(\frac{\omega_2 R_{\alpha,\beta}}{\hbar v_F}\right)
    +
    \sin\left(k_F R_{\alpha,\beta}-\frac{\pi}{4}\right)
    \sin\left(\frac{\omega_2 R_{\alpha,\beta}}{\hbar v_F}\right)
    \right]
  .
\end{eqnarray}
We denote by $\langle \langle\,...\,\rangle\rangle$ an averaging over
the oscillations at the scale of the Fermi wave-length, which yields a
simple expression for the {\it modes} consisting of the spin-up
electron and spin-down hole that counter-propagate along Andreev tubes
in the 2D conductor:
\begin{eqnarray}
  \label{eq:averaging1}
  && \langle\langle
  \hat{g}_{\alpha,\beta}^{A,1,1}\left(R_{\alpha,\beta},\omega_1\right)
  \hat{g}_{\beta,\alpha}^{A,2,2}\left(R_{\alpha,\beta},\omega_2\right)
  \rangle\rangle=\\&=&-\frac{1}{2 W^2 (k_F R_{\alpha,\beta})} \left\{
  \cos\left(\frac{\omega_1 R_{\alpha,\beta}}{\hbar v_F}\right)
  \cos\left(\frac{\omega_2 R_{\alpha,\beta}}{\hbar v_F}\right) -
  \sin\left(\frac{\omega_1 R_{\alpha,\beta}}{\hbar v_F}\right)
  \sin\left(\frac{\omega_2 R_{\alpha,\beta}}{\hbar v_F}\right)
  \right\} \\ &=& -\frac{1}{2 W^2 (k_F R_{\alpha,\beta})}
  \cos\left(\frac{(\omega_1+\omega_2)R_{\alpha,\beta}}{\hbar
    v_F}\right) .
  \label{eq:averaging2}
\end{eqnarray}

\section{General arguments for the current}
\label{sec:general}

In this section of the SM, we introduce general symmetry arguments for the
current, that will later be specialized to the $Q$-quartets,
$O$-octets and $Q'$-quartets. We consider the spectral current through
the $a$-$\alpha$ link \cite{Caroli1,Caroli2,Cuevas,Averin,Averin-noise}:
\begin{equation}
  \label{eq:Ia}
  I_a(\omega)=
  \hat{\tau}_3 \left(\hat{\Sigma}_{a,\alpha} \hat{G}^{+,-}_{\alpha,a}(\omega,\omega)
  - \hat{\Sigma}_{\alpha,a} \hat{G}^{+,-}_{a,\alpha}(\omega,\omega) \right)
  ,
\end{equation}
where $\hat{\tau}_3$ is one of the Pauli matrices acting in Nambu, and
$\hat{G}^{+,-}$ is the Keldysh Green's function. Combining the
Dyson-Keldysh Eq.~(\ref{eq:Gpm}) to
\begin{equation}
  \label{eq:product}
\left(AB\right)^{+,-}=A^{+,-} B^A + A^R B^{+,-}
,
\end{equation}
allows transforming Eq.~(\ref{eq:Ia}) into
\begin{equation}
  \label{eq:current-gene}
  I_a(\omega)=2\frac{(\Sigma_a)^2}{W}
  \left[ \exp\left(i\varphi_a\right) \hat{G}^{+,-,2,1}_{\alpha,\alpha}
    - \exp\left(-i\varphi_a\right) \hat{G}^{+,-,1,2}_{\alpha,\alpha}\right]
  ,
\end{equation}
where we implemented the large-gap approximation.

We now establish nonperturbative Keldysh calculations that connect the
snake diagrams to what we call as the current susceptibility $\partial
I_a/\partial \mu_N$, where $\mu_N$
is the electrochemical potential.

The fully dressed advanced or retarded Green's function
$\hat{G}_{\alpha,\alpha}^{A/R}$ at the tight-binding site $\alpha$ is
calculated from the Dyson equations, and it is expressed as an
infinite sum of terms that all involve products of the alternating
$\hat{g}$ and $\hat{\Sigma}$-type elements:
\begin{equation}
  \label{eq:Dy}
  \hat{G}_{\alpha,\alpha}=\hat{g}_{\alpha,\alpha} +
  \hat{g}_{\alpha,N}\hat{\Sigma}_{N,S} \hat{g}_{S,S}
  \hat{\Sigma}_{S,N'} \hat{g}_{N',\alpha}+ \hat{g}_{\alpha,N}
  \hat{\Sigma}_{N,S} \hat{g}_{S,S} \hat{\Sigma}_{S,N'} \hat{g}_{N',N''}
  \hat{\Sigma}_{N'',S'} \hat{g}_{S',S'} \hat{\Sigma}_{S',N'''}
  \hat{g}_{N''',\alpha} + ...,
\end{equation}
where $N$, $N'$, $N''$ and $N'''$ generically denotes a tight-binding
site in the normal conductor, at the contacts with the superconducting
leads that are generically denoted by $S$ and $S'$ in
Eq.~(\ref{eq:Dy}). Each term on this equation is a product of the
alternating $\hat{\Sigma}$ and $g$, and they all start with
$\hat{g}_{\alpha,N}$ and end with $\hat{g}_{N,\alpha}$. We note that
two types of snake diagrams are generally possible:

(i) The $\delta \hat{K}^{(\Sigma)}$ snake diagram orbits start with
$\hat{\Sigma}_{\alpha,a}$ and end with $\hat{g}_{N,\alpha}$.

(ii) The $\delta \hat{K}^{(g)}$ snake diagram orbits start with
$\hat{g}_{\alpha,N}$ and end with $\hat{\Sigma}_{a,\alpha}$.

We now take into account that $\hat{g}^{A,11}_{\alpha,\alpha} =
\hat{g}^{A,22}_{\alpha,\alpha} = i\pi\rho^{(0)}_{\alpha,\alpha}$ and
$\hat{g}^{R,11}_{\alpha,\alpha} = \hat{g}^{R,22}_{\alpha,\alpha} =
-i\pi\rho^{(0)}_{\alpha,\alpha}$ are pure imaginary, where
$\rho^{(0)}$ is the bare local density of states of the 2D conductor, see
Eq.~(\ref{eq:rho-0}).  Additionally taking the even parity of the
number of nonlocal bare Green's functions in the 2D normal conductor $N$
leads to the identities
\begin{eqnarray}
\label{eq:sym1}
\delta\hat{K}_{\alpha,\alpha}^{A,(\Sigma),1,1} &=&
\left(\delta\hat{K}_{\alpha,\alpha}^{R,(g),1,1}\right)^*\\
\delta\hat{K}_{\alpha,\alpha}^{A,(\Sigma),2,2} &=&
\left(\delta\hat{K}_{\alpha,\alpha}^{R,(g),2,2}\right)^*
.
\label{eq:sym2}
\end{eqnarray}
We additionally used that
\begin{equation}
  \label{eq:sym3}
\delta \hat{K}^{+,-,(\Sigma),1,1}_{\alpha,\alpha}
=-\left(\delta \hat{K}^{+,-,(g),1,1}_{\alpha,\alpha}\right)^*.
\end{equation}
Namely, the $(\Sigma)$ and the $(g)$ orbits run in opposite directions
and an odd number of (retarded, advanced) Green's functions have to
change into their (advanced, retarded) counterparts.

We now consider the connection between Eq.~(\ref{eq:A1}) for the
current, and the $\delta \hat{K}$ orbits discussed above. For this
purpose, let us consider changing $\hat{\Sigma}_{a,\alpha}^{1,1}
\hat{G}^{+,-,1,1}_{\alpha,a}$ into $\hat{\Sigma}_{\alpha,a}^{1,1}
\hat{G}^{+,-,1,1}_{a,\alpha}$ in Eq.~(\ref{eq:Ia}). It turns out that
the superconducting phase variables change sign in this
transformation, and we also note that, at the lowest order in
tunneling, each orbit has even length, and thus, an odd number of the
bare Green's functions has to change from advanced to retarded, or
from retarded to advanced. We deduce the following identity:
\begin{eqnarray}
  \label{eq:iden1}
\hat{\Sigma}_{a,\alpha}^{1,1} \hat{G}^{+,-,1,1}_{\alpha,a}-
\hat{\Sigma}_{\alpha,a}^{1,1} \hat{G}^{+,-,1,1}_{a,\alpha} &=& \delta
\hat{K}^{+,-,(g),1,1}-\delta\hat{K}^{+,-,(\Sigma),1,1}\\ &=&
-\left(\delta
\hat{K}^{+,-,(\Sigma),1,1}_{\alpha,\alpha}(\omega,\omega)\right)^*
-\delta \hat{K}^{+,-,(\Sigma),1,1}_{\alpha,\alpha}(\omega,\omega)
.
\label{eq:iden2}
\end{eqnarray}
We deduce from Eq.~(\ref{eq:A1}),
Eqs.~(\ref{eq:iden1})-(\ref{eq:iden2}), and a similar identity for the
$2,2$-Nambu component, the following expression of the current
oscillations $\delta I(\omega)$ associated to the $\delta
\hat{K}(\omega,\omega)$ orbit:
\begin{equation}
  \label{eq:I-om}
  \delta I(\omega)= -2 \mbox{Re}\left[ \delta
    \hat{K}^{+,-,(\Sigma),1,1}_{\alpha,\alpha}(\omega,\omega) - \delta
    \hat{K}^{+,-,(\Sigma),2,2}_{\alpha,\alpha}(\omega,\omega)\right]
  .
\end{equation}
We obtain the final expression of the spectral current associated to the
snake diagrams:
\begin{equation}
  \label{eq:current-v11}
  \delta I(\omega)=-2\left[ v_{1,1}(\omega,V,\mu_N)
    - v_{1,1}(\omega,-V,-\mu_N)\right]\cos\Psi
  ,
\end{equation}
where $v_{1,1}$ is defined as
\begin{eqnarray}
  \label{eq:v11-def}
  \delta \hat{K}_{\alpha,\alpha}^{+,-,(\Sigma/g),1,1}(\omega,\omega,V,\mu_N)&=& \pm
  v_{1,1}(\omega,V,\mu_N) \exp\left(\pm i \Psi\right)\\
  \delta \hat{K}_{\alpha,\alpha}^{+,-,(\Sigma/g),2,2}(\omega,\omega,V,\mu_N)&=& \pm
  v_{1,1}(\omega,-V,-\mu_N) \exp\left(\mp i \Psi\right)
  \label{eq:v22-def}
  .
\end{eqnarray}

\section{Two-terminal DC-Josephson effect with $\mu_N=0$ and $V=0$}
\label{sec:2T-eq}

In this section of the SM, we start with the DC-Josephson current between the
superconductors $S_L$ and $S_R$ connected at the tight-binding sites
$\alpha$ and $\beta$ on an infinite 2D conductor. We specifically consider
that a 2D rectangular normal region $N$ with $\mu_N=0$ is connected to
the two grounded superconducting leads $S_L$ and $S_R$ with
$V_L=V_R=0$, biased at the superconducting phases $\varphi_L$ and
$\varphi_R$. We denote by ``a'' and ``b'' the tight-binding sites of
$S_L$ and $S_R$ that make the contacts.

We deduce from Eqs.~(\ref{eq:I-omega-1})-(\ref{eq:I-omega-2}) the
following expression for the spectral current:
\begin{equation}
  \label{eq:integral-2T}
\int {\cal I}(\omega) d\omega = \frac{2i\Sigma_a^2\Sigma_b^2}{W^4(k_F
  R_{\alpha,\beta})} \sin(\varphi_R-\varphi_L) \int
n_F(\omega) \cos\left(\frac{2\omega R_{\alpha,\beta}}{\hbar
  v_F}\right) \frac{\Delta^2}{\Delta^2-(\omega-i\eta)^2}
d\omega
,
\end{equation}
where $\Sigma_{a,b}$ are the hopping amplitudes at the left and right
$(a,\alpha)$ and $(b,\beta)$ contacts, respectively, and $\Delta$ is
the superconducting gap. We deduce the low-energy $|\omega|\ll\Delta$
dimensionless two-terminal spectral supercurrent from
Eq.~(\ref{eq:integral-2T}):
\begin{equation}
i_{2T}(\omega) = \frac{1}{k_F R_{\alpha,\beta}}
\cos\left(\frac{2\omega R_{\alpha,\beta}}{\hbar
  v_F}\right),
\end{equation}
where we discarded the nonessential prefactors.

\section{$Q$-quartet channel}
\label{sec:Q}
In this section of the SM, we expand the $Q$-quartet current in
perturbation in the tunneling amplitudes and demonstrate Eq.~(3) in
the main text. Subsection~\ref{subsec1} evaluates the quartet current,
starting from Eq.~(\ref{eq:current-v11}). Subsection~\ref{subsec2}
evaluates a direct calculation of the quartet current, in agreement
with the results of subsection~\ref{subsec1}.

\subsection{Calculation of the quartet current
  from Eq.~(\ref{eq:current-v11})}
\label{subsec1}

In this subsection of the SM, we present a first argument for the quartet
current, deduced from the above
Eq.~(\ref{eq:current-v11}). Subsubsection~\ref{subsubsec1} evaluates
the snake diagram orbits of the
$Q$-quartets. Subsubsections~\ref{subsubsec2} and~\ref{subsubsec3}
evaluate the quartet current at equilibrium and at nonequilibrium,
respectively.

\subsubsection{Snake diagram orbits for the $Q$-quartets}
\label{subsubsec1}

In this subsubsection of the SM, we evaluate the $\delta
\hat{K}_{\alpha,\alpha}^{R,(\Sigma),1,1}$ quartet snake diagram
orbits:
\begin{eqnarray}
  \delta \hat{K}_{\alpha,\alpha}^{R,(\Sigma),1,1}(\omega,\omega)&=&
  \hat{\Sigma}_{\alpha,a}^{1,1}(\omega,\omega+eV) \hat{g}^{R,1,2}_{a,a}(\omega+eV,\omega+eV)
  \hat{\Sigma}_{a,\alpha}^{2,2}(\omega+eV,\omega+2eV)
  \hat{g}_{\alpha,\gamma}^{R,2,2}(\omega+2eV,\omega+2eV) \times\\\nonumber
&&  \hat{\Sigma}_{\gamma,c}^{2,2}(\omega+2eV,\omega+2eV)
  \hat{g}_{c,c}^{R,2,1}(\omega+2eV,\omega+2eV) \hat{\Sigma}_{c,\gamma}^{1,1}(\omega+2eV,\omega+2eV)
  \hat{g}^{R,1,1}_{\gamma,\beta}(\omega+2eV,\omega+2eV) \times\\\nonumber&&\hat{\Sigma}_{\beta,b}^{1,1}(\omega+2eV,\omega+eV)
  \hat{g}_{b,b}^{R,1,2}(\omega+eV,\omega+eV) \hat{\Sigma}_{b,\beta}^{2,2}(\omega+eV,\omega)
  \hat{g}_{\beta,\gamma}^{R,2,2}(\omega,\omega) \hat{\Sigma}_{\gamma,c}^{2,2}(\omega,\omega)
  \hat{g}_{c,c}^{R,2,1}(\omega,\omega) \times\\\nonumber&&\hat{\Sigma}_{c,\gamma}^{1,1}(\omega,\omega)
  \hat{g}_{\gamma,\alpha}^{R,1,1}(\omega,\omega)
  ,
\end{eqnarray}
and  
\begin{eqnarray}
  \delta \hat{K}_{\alpha,\alpha}^{R,(g),1,1}(\omega,\omega)&=&
  \hat{g}_{\alpha,\gamma}^{R,1,1}(\omega,\omega)
  \hat{\Sigma}_{\gamma,c}^{1,1}(\omega,\omega)
  \hat{g}_{c,c}^{R,1,2}(\omega,\omega)
  \hat{\Sigma}_{c,\gamma}^{2,2}(\omega,\omega)
  \hat{g}_{\gamma,\beta}^{R,2,2}(\omega,\omega)
  \hat{\Sigma}_{\beta,b}^{2,2}(\omega,\omega+eV)
  \hat{g}_{b,b}^{R,2,1}(\omega+eV,\omega+eV)\times\\\nonumber
&&  \hat{\Sigma}_{b,\beta}^{1,1}(\omega+eV,\omega+2eV)
  \hat{g}_{\beta,\gamma}^{R,1,1}(\omega+2eV,\omega+2eV)
  \hat{\Sigma}_{\gamma,c}^{1,1}(\omega+2eV,\omega+2eV)
  \hat{g}_{c,c}^{R,1,2}(\omega+2eV,\omega+2eV)\times\\\nonumber
&&  \hat{\Sigma}_{c,\gamma}^{2,2}(\omega+2eV,\omega+2eV)
  \hat{g}_{\gamma,\alpha}^{R,2,2}(\omega+2eV,\omega+2eV)
  \hat{\Sigma}_{\alpha,a}^{2,2}(\omega+2eV,\omega+eV) \times\\\nonumber
&&  \hat{g}_{a,a}^{R,2,1}(\omega+eV,\omega+eV)
  \hat{\Sigma}_{a,\alpha}^{1,1}(\omega,\omega)
  .
\end{eqnarray}

The Keldysh components $\delta
\hat{K}_{\alpha,\alpha}^{+,-,(\Sigma),1,1}$ and $\delta
\hat{K}_{\alpha,\alpha}^{+,-,(g),1,1}$ take the following form at the
lowest order in tunneling:
\begin{eqnarray}
&&  \delta \hat{K}_{\alpha,\alpha}^{+,-,(\Sigma),1,1}=
\frac{(\Sigma_L)^2 (\Sigma_R)^2 (\Sigma_B)^4}
     {W^4}
     \exp (i \Psi_Q)\times\\\nonumber
&&     \left\{\overline{\hat{g}^{+,-,2,2}_{\alpha,\gamma}(\omega+2eV) \hat{g}^{A,1,1}_{\gamma,\alpha}(\omega)}
     \times
     \overline{\hat{g}^{A,1,1}_{\gamma,\beta}(\omega+2eV) \hat{g}^{A,2,2}_{\beta,\gamma}(\omega)}+
     \overline{\hat{g}^{R,2,2}_{\alpha,\gamma}(\omega+2eV) \hat{g}^{A,1,1}_{\gamma,\alpha}(\omega)}
     \times
     \overline{\hat{g}^{+,-,1,1}_{\gamma,\beta}(\omega+2eV) \hat{g}^{A,2,2}_{\beta,\gamma}(\omega)}\right.
     \\\nonumber&&+\left.
     \overline{\hat{g}^{R,2,2}_{\alpha,\gamma}(\omega+2eV) \hat{g}^{A,1,1}_{\gamma,\alpha}(\omega)}
     \times
     \overline{\hat{g}^{R,1,1}_{\gamma,\beta}(\omega+2eV) \hat{g}^{+,-,2,2}_{\beta,\gamma}(\omega)}+
     \overline{\hat{g}^{R,2,2}_{\alpha,\gamma}(\omega+2eV) \hat{g}^{+,-,1,1}_{\gamma,\alpha}(\omega)}
     \times
     \overline{\hat{g}^{R,1,1}_{\gamma,\beta}(\omega+2eV) \hat{g}^{R,2,2}_{\beta,\gamma}(\omega)}
     \right\}
     ,
\end{eqnarray}
where the quartet phase is denoted by
$\Psi_Q=\varphi_L+\varphi_R-2\varphi_B$, and
\begin{eqnarray}
&&  \delta \hat{K}_{\alpha,\alpha}^{+,-,(g),1,1}=
\frac{(\Sigma_L)^2 (\Sigma_R)^2 (\Sigma_B)^4}
     {W^4}
     \exp\left(-i\Psi_Q\right)\times\\\nonumber
&&     \left\{\overline{\hat{g}^{+,-,1,1}_{\alpha,\gamma}(\omega) \hat{g}^{A,2,2}_{\gamma,\alpha}(\omega+2eV)}
     \times
     \overline{\hat{g}^{A,2,2}_{\gamma,\beta}(\omega) \hat{g}^{A,1,1}_{\beta,\gamma}(\omega+2eV)}+
     \overline{\hat{g}^{R,1,1}_{\alpha,\gamma}(\omega) \hat{g}^{A,2,2}_{\gamma,\alpha}(\omega+2eV)}
     \times
     \overline{\hat{g}^{+,-,2,2}_{\gamma,\beta}(\omega) \hat{g}^{A,1,1}_{\beta,\gamma}(\omega+2eV)}\right.
     \\\nonumber&&+\left.
     \overline{\hat{g}^{R,1,1}_{\alpha,\gamma}(\omega) \hat{g}^{A,2,2}_{\gamma,\alpha}(\omega+2eV)}
     \times
     \overline{\hat{g}^{R,2,2}_{\gamma,\beta}(\omega) \hat{g}^{+,-,1,1}_{\beta,\gamma}(\omega+2eV)}+
     \overline{\hat{g}^{R,1,1}_{\alpha,\gamma}(\omega) \hat{g}^{+,-,2,2}_{\gamma,\alpha}(\omega+2eV)}
     \times
     \overline{\hat{g}^{R,2,2}_{\gamma,\beta}(\omega) \hat{g}^{R,1,1}_{\beta,\gamma}(\omega+2eV)}
     \right\}
     .
\end{eqnarray}

Using Eqs.~(\ref{eq:averaging1})-(\ref{eq:averaging2}) to
evaluate the products of pairs of Green's functions leads to
the following expression of $v_{1,1}^{(Q)}$ defined in the above
Eq.~(\ref{eq:v11-def}):
\begin{eqnarray}
  v_{1,1}^{(Q)}(\omega,V,\mu_N)&=& - \frac{(\Sigma_L)^2 (\Sigma_R)^2
    (\Sigma_B)^4} {4 W^8 (k_F R_{\alpha,\gamma})(k_F
    R_{\gamma,\beta})}
  \cos\left(\frac{2(\omega+eV)R_{\alpha,\gamma}}{\hbar v_F}\right)
  \cos\left(\frac{2(\omega+eV)R_{\gamma,\beta}}{\hbar v_F}\right)
  \times \\\nonumber&&\left\{ n_F(\omega-\mu_N)-n_F(\omega+\mu_N)
  +n_F(\omega+2eV-\mu_N)-n_F(\omega+2eV+\mu_N) \right\}
  .
\end{eqnarray}

\subsubsection{Calculation of the $Q$-quartet spectral supercurrent
  at $\mu_N=0$ and $V=0$}
\label{subsubsec2}

In this subsubsection of the SM, we now extend the above section~\ref{sec:2T-eq}
to the ${Q}$-quartets. We deduce the dimensionless ${Q}$-quartet
spectral supercurrent:
\begin{equation}
  \label{eq:iQ-omega}
i_{Q}(\omega) = \frac{1}{(k_F R_{\alpha,\gamma}) (k_F
  R_{\gamma,\beta})} \cos\left(\frac{2\omega R_{\alpha,\gamma}}{\hbar
  v_F}\right) \cos\left(\frac{2\omega R_{\gamma,\beta}}{\hbar
  v_F}\right) ,
\end{equation}
where the notations $\alpha$, $\beta$ and $\gamma$ are provided in the
main text, see Figure 3a in the main text. This demonstrates Eq.~(1)
in the main text.

\subsubsection{Calculation of the $Q$-quartet current susceptibility
  with $\mu_N\ne 0$ and $V \ne 0$}
\label{subsubsec3}

In this subsubsection of the SM, we evaluate the $Q$-quartet current
susceptibility, which demonstrates Eq.~(3) in the main text. We obtain
the following expression for the quartet current:
\begin{eqnarray}
  \label{eq:IQ-om1}
&&I^{(Q)}(\omega) =
  \frac{(\Sigma_L)^2 (\Sigma_R)^2
    (\Sigma_B)^4}{2 W^8(k_F R_{\alpha,\gamma}) (k_F R_{\gamma,\beta})}
  \times \cos\Psi_Q \times\\ \nonumber && \left\{\left[n_F(\omega-\mu_N)
    -n_F(\omega+\mu_N)+n_F(\omega+2eV-\mu_N) - n_F(\omega+2eV+\mu_N)\right]
  \times
  \cos\left(\frac{2(\omega+eV) R_{\alpha,\gamma}}{\hbar v_F}\right)
  \cos\left(\frac{2(\omega+eV) R_{\gamma,\beta}}{\hbar v_F}\right) \right.\\&& -
  \nonumber
  \left.
  \left[n_F(\omega+\mu_N)
    -n_F(\omega-\mu_N)+n_F(\omega-2eV+\mu_N) - n_F(\omega-2eV-\mu_N)\right]
  \times
  \cos\left(\frac{2(\omega-eV) R_{\alpha,\gamma}}{\hbar v_F}\right)
  \cos\left(\frac{2(\omega-eV) R_{\gamma,\beta}}{\hbar v_F}\right)
  \right\}
  .
\end{eqnarray}
The corresponding current susceptibility is obtained by
differentiating Eq.~(\ref{eq:IQ-om1}) with respect to the
electrochemical potential $\mu_N$, and integrating over the energy
$\omega$:
\begin{equation}
  \label{eq:chiIQ}
  \chi_I^{(Q)}=\frac{\partial I^{(Q)}(\mu_N,eV)}{\partial\mu_N}=\int
  \frac{\partial}{\partial \mu_N} I^{(Q)}(\omega,\mu_N,eV) d\omega
  .
\end{equation}
Combining Eq.~(\ref{eq:IQ-om1}) to Eq.~(\ref{eq:chiIQ}) leads to
\begin{eqnarray}
  \label{eq:chiIQ2}
  \chi_I^{(Q)}&=& 2 \frac{(\Sigma_L)^2 (\Sigma_R)^2 (\Sigma_B)^4}{
    W^8 (k_F R_{\alpha,\gamma}) (k_F R_{\gamma,\beta})}
\times\cos\Psi_Q\times\\ &&\left\{ \cos\left(\frac{(2(\mu_N+eV)
      R_{\alpha,\gamma}}{\hbar v_F}\right)
    \cos\left(\frac{2(\mu_N+eV) R_{\gamma,\beta}}{\hbar v_F}\right) +
    \cos\left(\frac{2(\mu_N-eV) R_{\alpha,\gamma}}{\hbar v_F}\right)
    \cos\left(\frac{2(\mu_N-eV) R_{\gamma,\beta}}{\hbar
      v_F}\right)\right\}
    \nonumber
    ,
\end{eqnarray}
see Eq.~(3) in the main text.  {The dimensionless
  $Q$-quartet current susceptibility then takes the following form:
\begin{eqnarray}
  \label{eq:i'Q}
\frac{\partial i'_Q(eV,\mu_N)}{\partial \mu_N}
&=&
\frac{1}{(k_F R_{\alpha,\gamma}) (k_F R_{\gamma,\beta})} \left\{
\cos\left(\frac{2(\mu_N+eV) R_{\alpha,\gamma}}{\hbar v_F}\right)
\cos\left(\frac{2(\mu_N+eV) R_{\gamma,\beta}}{\hbar v_F}\right)\right.\\
&+& \left.
    \cos\left(\frac{2(\mu_N-eV) R_{\alpha,\gamma}}{\hbar v_F}\right)
    \cos\left(\frac{2(\mu_N-eV) R_{\gamma,\beta}}{\hbar
      v_F}\right)\right\}
    .
\nonumber
\end{eqnarray}
Eq.~(3) in the main text relates $\partial i'_Q(eV,\mu_N)/\partial
\mu_N$ in Eq.~(\ref{eq:i'Q}) to $i_Q(\omega)$ in
Eq.~(\ref{eq:iQ-omega}).  }

\subsection{Direct calculation of the quartet current}
\label{subsec2}

In this subsection of the SM, we recalculate the quartet current from a direct
evaluation.

\subsubsection{Expression for one of the components of the
  $Q$-quartet spectral supercurrent}

In this subsubsection of the SM, we present the first steps in a direct calculation of the
$Q$-quartet supercurrent. Specifically, we expand ${\cal I}_{Q}^{(A)}(\omega)= \left\{
\hat{\Sigma}_{a,\alpha} \hat{G}^{1,1}_{\alpha,a}\right\}^{+,-}$
according to
\begin{eqnarray}
  \label{eq:0C-1}
{\cal I}_{Q}^{(A)}(\omega)  &=&\left\{\hat{\Sigma}_{a,\alpha}^{1,1}(\omega,\omega-eV)
  \hat{g}^{1,1}_{\alpha,\gamma}(\omega-eV,\omega-eV)
  \hat{\Sigma}_{\gamma,c}^{1,1}(\omega-eV,\omega-eV)
  \hat{g}_{c,c}^{1,2}(\omega-eV,\omega-eV)
  \hat{\Sigma}_{c,\gamma}^{2,2}(\omega-eV,\omega-eV)\right.\\
  \nonumber
  &&
  \hat{g}_{\gamma,\beta}^{2,2}(\omega-eV,\omega-eV)
  \hat{\Sigma}_{\beta,b}^{2,2}(\omega-eV,\omega)
  \hat{g}_{b,b}^{2,1}(\omega,\omega)
  \hat{\Sigma}_{b,\beta}^{1,1}(\omega,\omega+eV)
  \hat{g}_{\beta,\gamma}^{1,1}(\omega+eV,\omega+eV)
  \hat{\Sigma}_{\gamma,c}^{1,1}(\omega+eV,\omega+eV)\\
  \nonumber
  &&\left.
  \hat{g}_{c,c}^{1,2}(\omega+eV,\omega+eV)
  \hat{\Sigma}_{c,\gamma}^{2,2}(\omega+eV,\omega+eV)
  \hat{g}_{\gamma,\alpha}^{2,2}(\omega+eV,\omega+eV)
  \hat{\Sigma}_{\alpha,a}^{2,2}(\omega+eV,\omega)
  \hat{g}_{a,a}^{2,1}(\omega,\omega)\right\}^{+,-}
  .
\end{eqnarray}
Gathering the Green's functions in a pair-wise manner leads to
\begin{eqnarray}
  \label{eq:0A-1}
{\cal I}_{Q}^{(A)}(\omega)  &=&
  \frac{\left(\Sigma_L\right)^2 \left(\Sigma_R\right)^2 \left(\Sigma_B\right)^4}{W^4}\times
    \exp\left[i\left(2\varphi_B-\varphi_L-\varphi_R\right)\right] \times
    \overline{\hat{g}^{1,1}_{\alpha,\gamma}(\omega-eV) \hat{g}^{2,2}_{\gamma,\alpha}(\omega+eV)}\times
    \overline{\hat{g}^{1,1}_{\beta,\gamma}(\omega+eV) \hat{g}^{2,2}_{\gamma,\beta}(\omega-eV)}
    ,
\end{eqnarray}
where the Keldysh component is implicit in this
Eq.~(\ref{eq:0A-1}). Now making the Keldysh component explicit leads
to
\begin{eqnarray}
  \label{eq:0-1}
  {\cal I}_{Q}^{(A)}(\omega)  &=& 2
  \frac{\left(\Sigma_L\right)^2 \left(\Sigma_R\right)^2 \left(\Sigma_B\right)^4}{W^4}\times
  \exp\left[i\left(2\varphi_B-\varphi_L-\varphi_R\right)\right]\times
  \overline{\hat{g}^{1,1,A}_{\alpha,\gamma}(\omega-eV) \hat{g}^{2,2,A}_{\gamma,\alpha}(\omega+eV)}\times
  \overline{\hat{g}^{1,1,A}_{\beta,\gamma}(\omega+eV) \hat{g}^{2,2,A}_{\gamma,\beta}(\omega-eV)}\times\\
  \nonumber
  &&
  \left\{
  n_F(\omega-eV-\mu_N) - n_F(\omega-eV+\mu_N) + n_F(\omega+eV-\mu_N) - n_F(\omega+eV+\mu_N)
  \right\}
  ,
\end{eqnarray}
where we used
$\hat{g}^A_{\alpha_k,\alpha_l}=-\hat{g}^R_{\alpha_k,\alpha_l}$ in the
infinite 2D conductor. The factor of $2$ in Eq.~(\ref{eq:0-1})
originates from the following expression of the 2D conductor bare
Keldysh Green's function: $\hat{g}^{+,-}_{\alpha_k,\alpha_l} = n_F
(\hat{g}^A_{\alpha_k,\alpha_l} - \hat{g}^R_{\alpha_k,\alpha_l}) = 2
n_F \hat{g}^A_{\alpha_k,\alpha_l}$. Taking the derivative with respect
to the electrochemical potential $\mu_N$ leads to
\begin{eqnarray}
  \label{eq:0B-1}
\frac{\partial  {\cal I}_{Q}^{(A)}(\omega)}{\partial \mu_N}&=& 2
\frac{\left(\Sigma_L\right)^2 \left(\Sigma_R\right)^2
  \left(\Sigma_B\right)^4}{W^4}\times
\exp\left[i\left(2\varphi_B-\varphi_L-\varphi_R\right)\right]\times
\overline{\hat{g}^{1,1,A}_{\alpha,\gamma}(\omega-eV)
  \hat{g}^{2,2,A}_{\gamma,\alpha}(\omega+eV)}\times
\overline{\hat{g}^{1,1,A}_{\beta,\gamma}(\omega+eV)
  \hat{g}^{2,2,A}_{\gamma,\beta}(\omega-eV)}\times\\ && \left\{
\delta(\omega-eV-\mu_N) +\delta(\omega-eV+\mu_N) +
\delta(\omega+eV-\mu_N) + \delta(\omega+eV+\mu_N)\right\} .
\nonumber
\end{eqnarray}

\subsubsection{Performing the integral over energy}
In this subsubsection of the SM, we proceed further by integrating
Eq.~(\ref{eq:0B-1}) over the energy $\omega$, which leads to the
following expression for the corresponding contribution to the current
susceptibility $I_{Q}^{(A)}=\int {\cal I}_{Q}^{(A)}(\omega) d \omega$:
\begin{eqnarray}
  \label{eq:pref-unity}
\frac{\partial I_{Q}^{(A)}}{\partial \mu_N} &=&
  \frac{\left(\Sigma_L\right)^2 \left(\Sigma_R\right)^2
    \left(\Sigma_B\right)^4}{W^8}\times \frac{1}{(k_F R_{\alpha,\gamma})(k_F
    R_{\gamma,\beta})}
  \exp\left[i\left(2\varphi_B-\varphi_L-\varphi_R\right)\right]\times\\ &&
  \left\{\cos\left(\frac{(2eV+2\mu_N) R_{\alpha,\gamma}}{\hbar
    v_F}\right) \cos\left(\frac{(2eV+2\mu_N) R_{\gamma,\beta}}{\hbar
    v_F}\right) + \cos\left(\frac{(2eV-2\mu_N) R_{\alpha,\gamma}}{\hbar
    v_F}\right) \cos\left(\frac{(2eV-2\mu_N) R_{\gamma,\beta}}{\hbar
    v_F}\right)\right\} .
  \nonumber
\end{eqnarray}
In this equation, a prefactor of $1/4$ originates from averaging a pair
of transmission modes over the oscillations at the scale of the Fermi wave-length.
A factor of $2$ originates from $\hat{g}^{+,-}$ and another factor of $2$ originates
from the $\delta$-functions in energy. This overall produces a prefactor of unity in
Eq.~(\ref{eq:pref-unity}).

\subsubsection{Taking into account the other
  component of the spectral supercurrent}

In this subsubsection of the SM, we present the final step in the
evaluation of the quartet current susceptibility. The expression for
${\cal I}_{Q}^{(B)}(\omega)= \left\{ \hat{\Sigma}_{a,\alpha}
\hat{G}^{2,2}_{\alpha,a}\right\}^{+,-}$ is deduced from ${\cal
  I}_{Q}^{(A)}(\omega)$ in Eq.~(\ref{eq:0C-1}) by exchanging the
electron and the hole Nambu labels. As a result, the superconducting
phase variables and the electrochemical potential change sign in
Eq.~(\ref{eq:0-1}) when going from ${\cal I}_{Q}^{(A)}(\omega)$ to
${\cal I}_{Q}^{(B)}(\omega)$:
\begin{equation}
{\cal I}_{Q}^{(B)}\left(\omega,\{\varphi_n\},eV,\mu_N\right)
  =
  -
  {\cal I}_{Q}^{(A)}\left(\omega,\{-\varphi_n\},eV,\mu_N\right)
    .
\end{equation}
{We deduce the previous Eq.~(\ref{eq:chiIQ2}) for the
  quartet current susceptibility.}

\begin{figure*}[htb]
  \includegraphics[width=.49\textwidth]{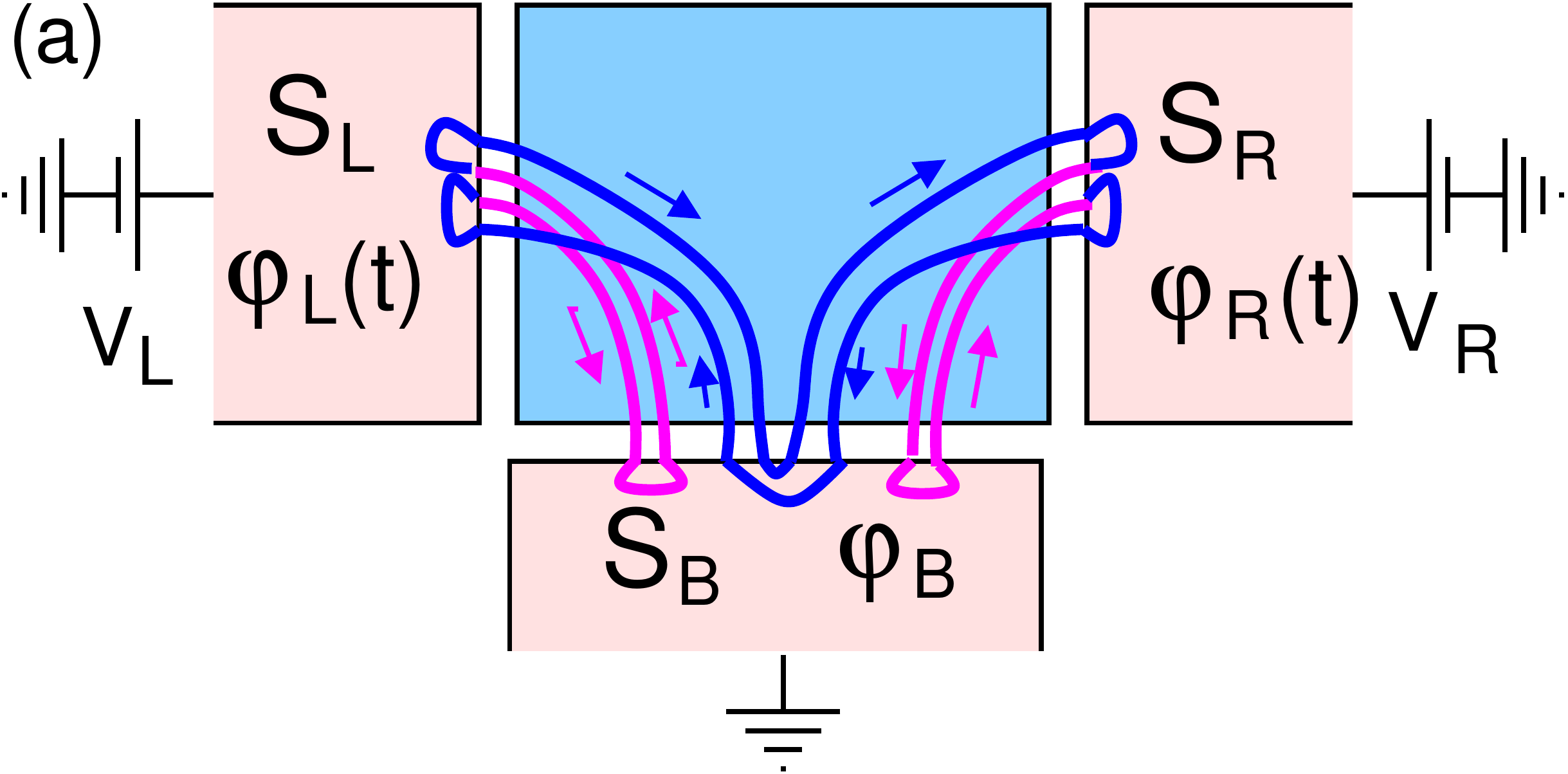}\includegraphics[width=.49\textwidth]{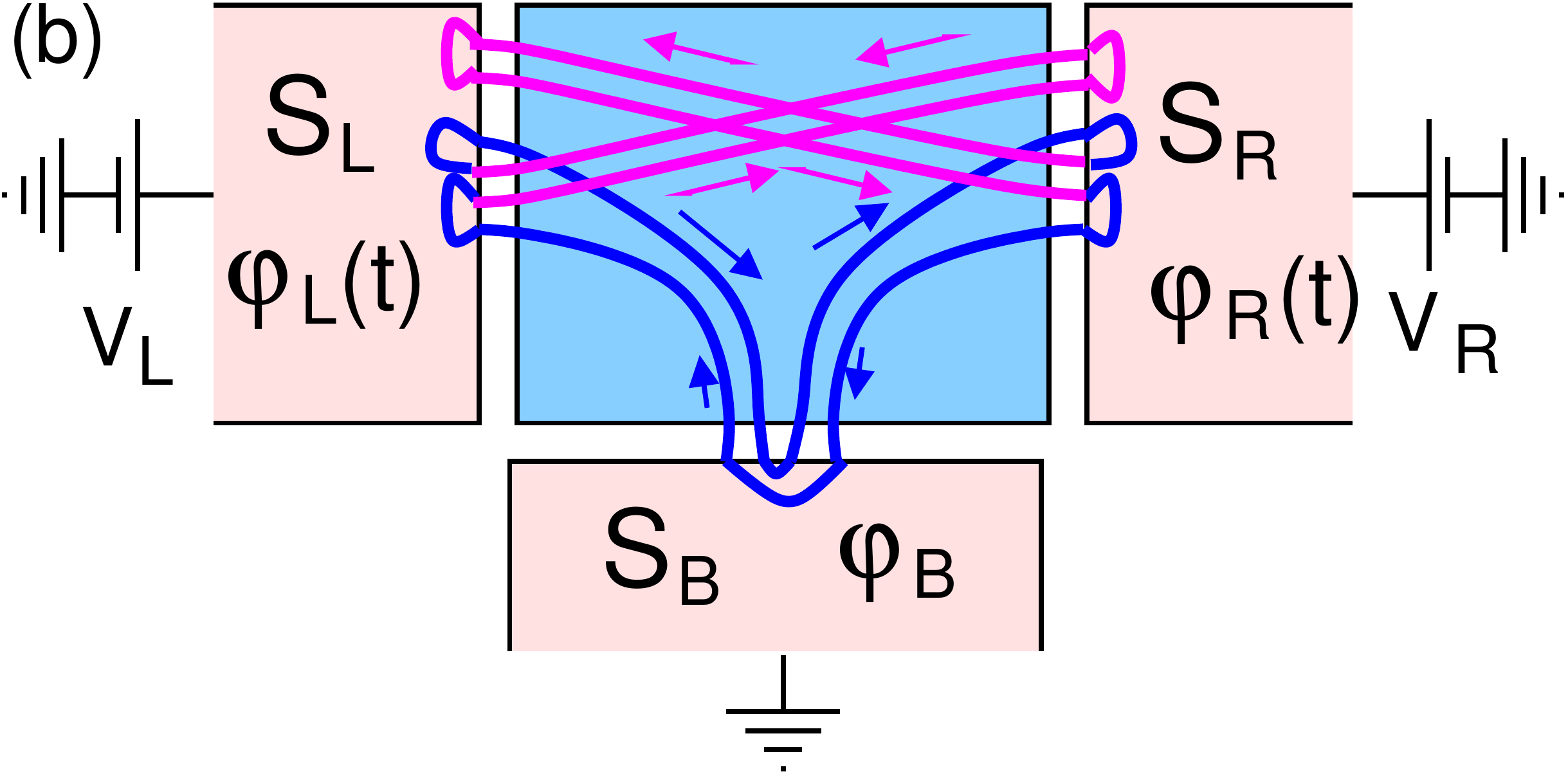}

  \caption{The $O$-octet and the $Q'$-quartet diagrams, see also
    Fig.~2c and e in the main text. \label{fig:O-Q'}}
\end{figure*}

\section{$O$-octet channel}
\label{sec:O}
In this section of the SM, we now address similar calculations for the
$O$-octets, see the corresponding diagram in Fig.\ref{fig:O-Q'}a of
the SM. The calculations presented in this section demonstrate Eq.~(4)
in the main text.  Subsection~\ref{subsecbis1} evaluates the $O$-octet
spectral current at equilibrium and
subsection~\ref{sec:higher-order-O} deals with the current
susceptibility at nonequilibrium.

\subsection{Demonstration of the higher-order ${O}$-octet
  spectral supercurrent with $\mu_N=0$ and $V=0$}
\label{subsecbis1}

In this subsection of the SM, we evaluate the higher-order $O$-octet spectral
supercurrent at equilibrium, i.e. with $\mu_N=V=0$. We obtain the
following dimensionless ${O}$-octet spectral supercurrent:
\begin{equation}
  \label{eq:iO-omega}
i_{O}(\omega) = \frac{1}{(k_F R_{\alpha,\gamma}) (k_F
  R_{\gamma,\beta}) (k_F R_{\alpha,\alpha'}) (k_F R_{\beta,\beta'})}
\cos\left(\frac{2\omega R_{\alpha,\gamma}}{\hbar v_F}\right)
\cos\left(\frac{2\omega R_{\gamma,\beta}}{\hbar v_F}\right)
\cos\left(\frac{2\omega R_{\alpha,\alpha'}}{\hbar v_F}\right)
\cos\left(\frac{2\omega R_{\beta,\beta'}}{\hbar v_F}\right)
,
\end{equation}
where we used the straightforward notations $\alpha$, $\alpha'$,
$\beta$, $\beta'$ and $\gamma$ to label the contact points of the
$O$-octet diagram. This demonstrates that Eq.~(1) in the main text
can still be used for the $Q$-quartets.

\subsection{Demonstration of the $O$-octet current susceptibility
  with $\mu_N\ne 0$ and $V \ne 0$}
\label{sec:higher-order-O}

In this subsection of the SM, we evaluate the $O$-octet current susceptibility
at finite bias. The expression of the $O$-octet spectral supercurrent
is presented in subsubsection~\ref{subsubtoto1}. The integral over
energy is presented in
subsubsection~\ref{sec:large-scale-interfaces1}. The final expression
of the $O$-octet supercurrent is presented in
subsubsection~\ref{subsubsectoto3}.

\subsubsection{Expression for one of the components of the $O$-octet spectral supercurrent}
\label{subsubtoto1}
In this subsubsection of the SM, we present the expression of the
$O$-octet spectral supercurrent. Similarly to the above calculations,
we expand ${\cal I}_{O}^{(A)}(\omega) = \left\{
\hat{\Sigma}_{a,\alpha} \hat{G}^{1,1}_{\alpha,a}\right\}^{+,-}$
according to
\begin{eqnarray}
{\cal I}_{O}^{(A)}(\omega)
    &=& \left\{\hat{\Sigma}_{a,\alpha}^{1,1}(\omega,\omega-eV)
    \hat{g}^{1,1}_{\alpha,\epsilon}(\omega-eV,\omega-eV)
    \hat{\Sigma}_{\epsilon,e}^{1,1}(\omega-eV,\omega-eV)
    \hat{g}_{e,e}^{1,2}(\omega-eV,\omega-eV)
    \hat{\Sigma}_{e,\epsilon}^{2,2}(\omega-eV,\omega-eV)\times\right.\\ &&
    \nonumber
    \hat{g}_{\epsilon,\beta}^{2,2}(\omega-eV,\omega-eV)
    \hat{\Sigma}_{\beta,b}^{2,2}(\omega-eV,\omega)
    \hat{g}_{b,b}^{2,1}(\omega,\omega)
    \hat{\Sigma}_{b,\beta}^{1,1}(\omega,\omega+eV)\times\\&&
    \nonumber
    \hat{g}_{\beta,\delta}^{1,1}(\omega+eV,\omega+eV)
    \hat{\Sigma}_{\delta,d}^{1,1}(\omega+eV,\omega+eV)
    \hat{g}_{d,d}^{1,2}(\omega+eV,\omega+eV)
    \hat{\Sigma}_{d,\delta}^{2,2}(\omega+eV,\omega+eV)\times\\&&
    \nonumber
    \hat{g}_{\delta,\beta}^{2,2}(\omega+eV,\omega+eV)
    \hat{\Sigma}_{\beta,b}^{2,2}(\omega+eV,\omega+2eV)
    \hat{g}_{b,b}^{2,1}(\omega+2eV,\omega+2eV)
    \hat{\Sigma}_{b,\beta}^{1,1}(\omega+2eV,\omega+3eV)\times\\&&
    \nonumber
    \hat{g}_{\beta,\epsilon}^{1,1}(\omega+3eV,\omega+3eV)
    \hat{\Sigma}_{\epsilon,e}^{1,1}(\omega+3eV,\omega+3eV)
    \hat{g}_{e,e}^{1,2}(\omega+3eV,\omega+3eV)
    \hat{\Sigma}_{e,\epsilon}^{2,2}(\omega+3eV,\omega+3eV)\times\\&&
    \nonumber
    \hat{g}_{\epsilon,\alpha}^{2,2}(\omega+3eV,\omega+3ev)
    \hat{\Sigma}_{\alpha,a}^{2,2}(\omega+3eV,\omega+2eV)
    \hat{g}_{a,a}^{2,1}(\omega+2eV,\omega+2eV)
    \hat{\Sigma}_{a,\alpha}^{1,1}(\omega+2eV,\omega+eV)\times\\ &&
    \nonumber
    \hat{g}_{\alpha,\gamma}^{1,1}(\omega+eV,\omega+eV)
    \hat{\Sigma}_{\gamma,c}^{1,1}(\omega+eV,\omega+eV)
    \hat{g}_{c,c}^{1,2}(\omega+eV,\omega+eV)
    \hat{\Sigma}_{c,\gamma}^{2,2}(\omega+eV,\omega+eV) \times\\&&
    \nonumber
    \left.
    \hat{g}_{\gamma,\alpha}^{2,2}(\omega+eV,\omega+eV)
    \hat{\Sigma}_{\alpha,a}^{2,2}(\omega+eV,\omega)
    \hat{g}_{a,a}^{2,1}(\omega,\omega)\right\}^{+,-} .
  \end{eqnarray}
The same calculation as above leads to
\begin{eqnarray}
  \label{eq:I-O-A}
&& \frac{\partial {\cal I}_{O}^{(A)}(\omega)}{\partial \mu_N} =
    \frac{\left(\Sigma_L\right)^4 \left(\Sigma_R\right)^4
      \left(\Sigma_B\right)^8}{8 W^{16}}\times
    \exp\left(-i\Psi_O\right)\times
    \frac{1}{(k_F R_{\alpha,\epsilon})(k_F R_{\beta,\epsilon}) (k_F
      R_{\beta,\delta}) (k_F R_{\alpha,\gamma})} \times\\&&\nonumber
    \cos\left[\frac{(2\omega+2eV) R_{\alpha,\epsilon}}{\hbar
        v_F}\right] \cos\left[\frac{(2\omega+2eV)
        R_{\beta,\epsilon}}{\hbar v_F}\right]
    \cos\left[\frac{(2\omega+2eV) R_{\beta,\delta}}{\hbar v_F}\right]
    \cos\left[\frac{(2\omega+2eV) R_{\alpha,\gamma}}{\hbar
        v_F}\right]\\\nonumber &&\times \left\{ \delta(\omega-eV-\mu_N) +
    \delta(\omega-eV+\mu_N) + 2 \delta(\omega+eV-\mu_N) + 2
    \delta(\omega+eV+\mu_N)\right.\\ \nonumber&&\left.  +
    \delta(\omega+3eV-\mu_N) + \delta(\omega+3eV+\mu_N)\right\} .
  \end{eqnarray}
  In this expression, the four transmission modes produce a $1/16$ prefactor, and
  the bare Keldysh Green's function produces a factor of $2$, resulting in the
  overall $1/8$ prefactor.
  
\subsubsection{Performing the integral over energy}
\label{sec:large-scale-interfaces1}

In this subsubsection of the SM, we proceed further by evaluating the
energy-integral of the $O$-octet spectral supercurrent. Specifically,
integrating Eq.~(\ref{eq:I-O-A}) over the energy $\omega$ leads to
\begin{eqnarray}
  \label{eq:I-O-A-integre}
    && \frac{\partial I_O^{(A)}}{\partial \mu_N}= \int \frac{\partial
      {\cal I}_{O}^{(A)}(\omega)}{\partial \mu_N} d\omega=
    \frac{\left(\Sigma_L\right)^4 \left(\Sigma_R\right)^4
      \left(\Sigma_B\right)^8}{8 W^{16}}\times
    \exp\left(-i\Psi_O\right)\times\\ \nonumber&& \frac{1}{(k_F
      R_{\alpha,\epsilon})(k_F R_{\beta,\epsilon}) (k_F
      R_{\beta,\delta}) (k_F R_{\alpha,\gamma})} \times \left[
      A_1+A_2+A_3+A_4+A_5+A_6 \right] ,
  \end{eqnarray}
  where $A_1,\,...,\,A_6$ are given by
  \begin{eqnarray}
    \label{eq:An-1}
    A_1 &=& A_6=
    \cos\left[\frac{(4eV+2\mu_N) R_{\alpha,\epsilon}}{\hbar v_F} \right]
    \cos\left[\frac{(4eV+2\mu_N) R_{\beta,\epsilon}}{\hbar v_F} \right]
    \cos\left[\frac{(4eV+2\mu_N) R_{\beta,\delta}}{\hbar v_F} \right]
    \cos\left[\frac{(4eV+2\mu_N) R_{\alpha,\gamma}}{\hbar v_F} \right]\\
    A_2 &=& A_5=
    \cos\left[\frac{(4eV-2\mu_N) R_{\alpha,\epsilon}}{\hbar v_F} \right]
    \cos\left[\frac{(4eV-2\mu_N) R_{\beta,\epsilon}}{\hbar v_F} \right]
    \cos\left[\frac{(4eV-2\mu_N) R_{\beta,\delta}}{\hbar v_F} \right]
    \cos\left[\frac{(4eV-2\mu_N) R_{\alpha,\gamma}}{\hbar v_F} \right]\\
    A_3 &=& A_4 = 2
    \cos\left[\frac{2\mu_N R_{\alpha,\epsilon}}{\hbar v_F} \right]
    \cos\left[\frac{2\mu_N R_{\beta,\epsilon}}{\hbar v_F} \right]
    \cos\left[\frac{2\mu_N R_{\beta,\delta}}{\hbar v_F} \right]
    \cos\left[\frac{2\mu_N R_{\alpha,\gamma}}{\hbar v_F} \right]
    \label{eq:An-2}
    ,
\end{eqnarray}
  where the octet phase is defined as $\Psi_O=2\varphi_L+2\varphi_R-4\varphi_B$.
  
\subsubsection{Final expression for the $\cos \Psi_O$-component of the ${O}$-octet supercurrent}
\label{subsubsectoto3}

In this subsubsection of the SM, we conclude the calculation of the
$O$-octet supercurrent. Taking all components of the supercurrent
into account leads to the following final form of the partial
derivative of the ${O}$-octet supercurrent with respect to the
electrochemical potential $\mu_N$:
\begin{eqnarray}
  \label{eq:result3}
  \frac{\partial I_{O}}{\partial \mu_N} &=&
  \frac{\left(\Sigma_L\right)^4 \left(\Sigma_R\right)^4
    \left(\Sigma_B\right)^8}{2 W^{16}} \times \cos\Psi_O\times
  \frac{1}{(k_F R_{\alpha,\epsilon})(k_F R_{\beta,\epsilon}) (k_F
    R_{\beta,\delta}) (k_F R_{\alpha,\gamma})}\times\\ \nonumber&& \left\{
  \cos\left[\frac{(4eV+2\mu_N) R_{\alpha,\epsilon}}{\hbar v_F} \right]
  \cos\left[\frac{(4eV+2\mu_N) R_{\beta,\epsilon}}{\hbar v_F} \right]
  \cos\left[\frac{(4eV+2\mu_N) R_{\beta,\delta}}{\hbar v_F} \right]
  \cos\left[\frac{(4eV+2\mu_N) R_{\alpha,\gamma}}{\hbar v_F}
    \right]\right.\\\nonumber &&+ \cos\left[\frac{(4eV-2\mu_N)
      R_{\alpha,\epsilon}}{\hbar v_F} \right]
  \cos\left[\frac{(4eV-2\mu_N) R_{\beta,\epsilon}}{\hbar v_F} \right]
  \cos\left[\frac{(4eV-2\mu_N) R_{\beta,\delta}}{\hbar v_F} \right]
  \cos\left[\frac{(4eV-2\mu_N) R_{\alpha,\gamma}}{\hbar v_F}
    \right]\\ \nonumber&&+ 2 \left.  \cos\left[\frac{2\mu_N
      R_{\alpha,\epsilon}}{\hbar v_F} \right] \cos\left[\frac{2\mu_N
      R_{\beta,\epsilon}}{\hbar v_F} \right] \cos\left[\frac{2\mu_N
      R_{\beta,\delta}}{\hbar v_F} \right] \cos\left[\frac{2\mu_N
      R_{\alpha,\gamma}}{\hbar v_F} \right]\right\} .
\end{eqnarray}
The $1/2$ prefactor in Eq.~(\ref{eq:result3}) results from the $1/8$
prefactor in Eq.~(\ref{eq:I-O-A-integre}), that is multiplied by a
factor of two for $\exp(i\Psi_O)+\exp(-i\Psi_0)=2\cos\Psi_O$ from both
terms contributing to the current, and by another factor of $2$ for
doubling of the $A_n$s in Eqs.~(\ref{eq:An-1})-(\ref{eq:An-2}).

{The $O$-octet current susceptibility $\partial i'_O/\partial \mu_N(eV,\mu_N)$
  is given by
  \begin{eqnarray}
  \label{eq:i'O}
  \frac{\partial i'_{O}(eV,\mu_N)}{\partial \mu_N} &=&
  \frac{1}{(k_F R_{\alpha,\epsilon})(k_F R_{\beta,\epsilon}) (k_F
    R_{\beta,\delta}) (k_F R_{\alpha,\gamma})}\times\\ \nonumber&& \left\{
  \cos\left[\frac{(4eV+2\mu_N) R_{\alpha,\epsilon}}{\hbar v_F} \right]
  \cos\left[\frac{(4eV+2\mu_N) R_{\beta,\epsilon}}{\hbar v_F} \right]
  \cos\left[\frac{(4eV+2\mu_N) R_{\beta,\delta}}{\hbar v_F} \right]
  \cos\left[\frac{(4eV+2\mu_N) R_{\alpha,\gamma}}{\hbar v_F}
    \right]\right.\\\nonumber &&+ \cos\left[\frac{(4eV-2\mu_N)
      R_{\alpha,\epsilon}}{\hbar v_F} \right]
  \cos\left[\frac{(4eV-2\mu_N) R_{\beta,\epsilon}}{\hbar v_F} \right]
  \cos\left[\frac{(4eV-2\mu_N) R_{\beta,\delta}}{\hbar v_F} \right]
  \cos\left[\frac{(4eV-2\mu_N) R_{\alpha,\gamma}}{\hbar v_F}
    \right]\\ \nonumber&&+ 2 \left.  \cos\left[\frac{2\mu_N
      R_{\alpha,\epsilon}}{\hbar v_F} \right] \cos\left[\frac{2\mu_N
      R_{\beta,\epsilon}}{\hbar v_F} \right] \cos\left[\frac{2\mu_N
      R_{\beta,\delta}}{\hbar v_F} \right] \cos\left[\frac{2\mu_N
      R_{\alpha,\gamma}}{\hbar v_F} \right]\right\} .
  \nonumber
\end{eqnarray}
  Eq.~(4) in the main text relates $\partial i'_O(eV,\mu_N)/\partial
  \mu_N$ in Eq.~(\ref{eq:i'O}) to $i_O$ in Eq.~(\ref{eq:iO-omega}).}

\section{$Q'$-quartet channel}
\label{sec:Q'}
In this section of the SM, we similarly treat the $Q'$-quartet
channel, see the corresponding diagram in Fig.~\ref{fig:O-Q'}b of the
SM, and demonstrate Eq.~(5) in the main text. The equilibrium
$Q'$-quartet spectral supercurrent is discussed in
subsection~\ref{subsec:titi1}. The nonequilibrium $Q'$-quartet
susceptibility is calculated in subsection~\ref{sec:higher-order-Q'}.

\subsection{Demonstration of the higher-order ${Q}'$-quartet
  spectral supercurrent with $\mu_N=0$ and $V=0$}
\label{subsec:titi1}

In this subsection of the SM, we calculate the $Q'$-quartet spectral
supercurrent at equilibrium, i.e. with $\mu_N=V=0$. We obtain the
following expression for the higher-order ${Q}'$-quartet diagram:
\begin{equation}
  \label{eq:iQ'-omega}
i_{{Q}'}(\omega) = \frac{1}{(k_F R_{\alpha,\gamma}) (k_F
  R_{\gamma,\beta}) (k_F R_{\alpha,\beta'}) (k_F R_{\beta,\alpha'})}
\cos\left(\frac{2\omega R_{\alpha,\gamma}}{\hbar v_F}\right)
\cos\left(\frac{2\omega R_{\gamma,\beta}}{\hbar v_F}\right)
\cos\left(\frac{2\omega R_{\alpha,\beta'}}{\hbar v_F}\right)
\cos\left(\frac{2\omega R_{\beta,\alpha'}}{\hbar v_F}\right) ,
\end{equation}
where the notations $\alpha$, $\alpha'$, $\beta$, $\beta'$ and
$\gamma$ are used to label the contact points, see Figure 3a in the
main text.

\subsection{Demonstration of the $Q'$-quartet current susceptibility
  with $\mu_N\ne 0$ and $V \ne 0$}

\label{sec:higher-order-Q'}

In this subsection of the SM, we evaluate the $Q'$-quartet susceptibility at
finite bias. One of the components of the $Q'$-quartet spectral
supercurrent is evaluated in subsubsection~\ref{subsubtata1}. The
integral over energy is evaluated in
subsubsection~\ref{sec:large-scale-interfaces2}. The final expression
of the $Q'$-quartet supercurrent is obtained in
subsubsection~\ref{subsubtata3}.

\subsubsection{Expression for one of the components of the $Q'$-quartet spectral supercurrent}
\label{subsubtata1}

In this subsubsection of the SM, we present the first steps in the
calculation of the $Q'$-quartet supercurrent. Similarly as above, we
find the following expression for ${\cal I}_{{Q}'}^{(A)}(\omega)=
\left\{ \hat{\Sigma}_{a,\alpha}
\hat{G}^{1,1}_{\alpha,a}\right\}^{+,-}$:
\begin{eqnarray}
{\cal I}_{{Q}'}^{(A)}(\omega) &=& \left\{\hat{\Sigma}_{\alpha,a}^{1,1}(\omega,\omega+eV)
    \hat{g}_{a,a}^{1,2}(\omega+eV,\omega+eV)
    \hat{\Sigma}_{a,\alpha}^{2,2}(\omega+eV,\omega+2eV)
    \hat{g}_{\alpha,\delta}^{2,2}(\omega+2eV,\omega+2eV)
    \hat{\Sigma}_{\delta,d}^{2,2}(\omega+2eV,\omega+3eV)\times\right.\\ &&
    \nonumber
    \hat{g}_{d,d}^{2,1}(\omega+3eV,\omega+3eV)
    \hat{\Sigma}_{d,\delta}^{1,1}(\omega+3eV,\omega+4eV)
    \hat{g}_{\delta,\epsilon}^{1,1}(\omega+4eV,\omega+4eV)
    \hat{\Sigma}_{\epsilon,e}^{1,1}(\omega+4eV,\omega+4eV)\times\\&&
    \nonumber
    \hat{g}^{1,2}_{e,e}(\omega+4eV,\omega+4eV)
    \hat{\Sigma}^{2,2}_{e,\epsilon}(\omega+4eV,\omega+4eV)
    \hat{g}^{2,2}_{\epsilon,\beta}(\omega+4eV,\omega+4eV)
    \hat{\Sigma}_{\beta,b}^{2,2}(\omega+4eV,\omega+3eV)\times\\&&
    \nonumber
    \hat{g}_{b,b}^{2,1}(\omega+3eV,\omega+3eV)
    \hat{\Sigma}_{b,\beta}^{1,1}(\omega+3eV,\omega+2eV)
    \hat{g}_{\beta,\gamma}^{1,1}(\omega+2eV,\omega+2eV)
    \hat{\Sigma}_{\gamma,c}^{1,1}(\omega+2eV,\omega+eV)\times\\&&
    \nonumber
    \hat{g}_{c,c}^{1,2}(\omega+eV,\omega+eV)
    \hat{\Sigma}_{c,\gamma}^{2,2}(\omega+eV,\omega)
    \hat{g}_{\gamma,\beta}^{2,2}(\omega,\omega)
    \hat{\Sigma}_{\beta,b}^{2,2}(\omega,\omega-eV)\times\\&&
    \nonumber
    \hat{g}_{b,b}^{2,1}(\omega-eV,\omega-eV)
    \hat{\Sigma}_{b,\beta}^{1,1}(\omega-eV,\omega-2eV)
    \hat{g}_{\beta,\epsilon}^{1,1}(\omega-2eV,\omega-2eV)
    \hat{\Sigma}_{\epsilon,e}^{1,1}(\omega-2eV,\omega-2eV)\times\\ &&
    \nonumber
    \hat{g}_{e,e}^{1,2}(\omega-2eV,\omega-2eV)
    \hat{\Sigma}_{e,\epsilon}^{2,2}(\omega-2eV,\omega-2eV)
    \hat{g}_{\epsilon,\delta}^{2,2}(\omega-2eV,\omega-2eV)
    \hat{\Sigma}_{\delta,d}^{2,2}(\omega-2eV,\omega-eV)\times\\&&
    \nonumber
    \left.
    \hat{g}_{d,d}^{2,1}(\omega-eV,\omega-eV)
    \hat{\Sigma}_{d,\delta}^{1,1}(\omega-eV,\omega)
    \hat{g}_{\delta,\alpha}^{1,1}(\omega,\omega)\right\}^{+,-}
    ,
  \end{eqnarray}
which leads to 
  \begin{eqnarray}
&& \frac{\partial {\cal I}_{{Q}'}^{(A)}(\omega)}{\partial \mu_N} =
    \frac{\left(\Sigma_L\right)^4 \left(\Sigma_R\right)^4
      \left(\Sigma_B\right)^8}{8 W^{16}}\times
    \exp\left(-i\Psi_Q\right)\times \frac{1}{(k_F
      R_{\alpha,\delta})(k_F R_{\delta,\epsilon}) (k_F
      R_{\beta,\epsilon}) (k_F R_{\gamma,\beta})} \times\\ \nonumber &&
    \cos\left[\frac{(2\omega+2eV) R_{\alpha,\delta}}{\hbar v_F}\right]
    \cos\left[\frac{(2\omega+2eV) R_{\delta,\epsilon}}{\hbar
        v_F}\right] \cos\left[\frac{(2\omega+2eV)
        R_{\beta,\epsilon}}{\hbar v_F}\right]
    \cos\left[\frac{(2\omega+2eV) R_{\gamma,\beta}}{\hbar
        v_F}\right]\times\\ \nonumber && \left\{ \delta(\omega+2eV+\mu_N) +
    \delta(\omega+4eV-\mu_N) + \delta(\omega+4eV+\mu_N) +
    \delta(\omega+2eV-\mu_N) + \delta(\omega+\mu_N) +
    \delta(\omega-2eV-\mu_N)\right.\\\nonumber && + \left.
    \delta(\omega-2eV+\mu_N) + \delta(\omega-\mu_N)\right\}
  \end{eqnarray}

\subsubsection{Performing the integral over energy}
\label{sec:large-scale-interfaces2}

In this subsubsection of the SM, we present the result of integrating
the $Q'$-quartet supercurrent over the energy $\omega$. Similarly as
above, we find the following for $I_{{Q}'}=\int {\cal
  I}_{{Q}'}(\omega) d\omega$:
  \begin{eqnarray}
    \frac{\partial I_{{Q}'}}{\partial \mu_N} &=&
    \frac{\left(\Sigma_L\right)^4 \left(\Sigma_R\right)^4
      \left(\Sigma_B\right)^8}{8 W^{16}} \times
    \exp\left(-i\Psi_Q\right)\times
    \frac{1}{(k_F R_{\alpha,\delta})(k_F R_{\delta,\epsilon}) (k_F
      R_{\beta,\epsilon}) (k_F R_{\gamma,\beta})} \times\\ \nonumber&& \left[
      B_1+B_2+B_3+B_4+B_5+B_6+B_7+B_8 \right] ,
  \end{eqnarray}
  where $B_1,\,...,\,B_8$ are given by
  \begin{eqnarray}
    B_1 &=& B_8 =
    \cos\left[\frac{(2eV+2\mu_N) R_{\alpha,\delta}}{\hbar v_F} \right]
    \cos\left[\frac{(2eV+2\mu_N) R_{\delta,\epsilon}}{\hbar v_F} \right]
    \cos\left[\frac{(2eV+2\mu_N) R_{\beta,\epsilon}}{\hbar v_F} \right]
    \cos\left[\frac{(2eV+2\mu_N) R_{\gamma,\beta}}{\hbar v_F} \right]\\
    B_2 &=& B_7 =
    \cos\left[\frac{(-6eV+2\mu_N) R_{\alpha,\delta}}{\hbar v_F} \right]
    \cos\left[\frac{(-6eV+2\mu_N) R_{\delta,\epsilon}}{\hbar v_F} \right]
    \cos\left[\frac{(-6eV+2\mu_N) R_{\beta,\epsilon}}{\hbar v_F} \right]
    \cos\left[\frac{(-6eV+2\mu_N) R_{\gamma,\beta}}{\hbar v_F} \right]\\
    B_3 &=& B_6 =
    \cos\left[\frac{(6eV+2\mu_N) R_{\alpha,\delta}}{\hbar v_F} \right]
    \cos\left[\frac{(6eV+2\mu_N) R_{\delta,\epsilon}}{\hbar v_F} \right]
    \cos\left[\frac{(6eV+2\mu_N) R_{\beta,\epsilon}}{\hbar v_F} \right]
    \cos\left[\frac{(6eV+2\mu_N) R_{\gamma,\beta}}{\hbar v_F} \right]\\
    B_4 &=& B_5 =
    \cos\left[\frac{(-2eV+2\mu_N) R_{\alpha,\delta}}{\hbar v_F} \right]
    \cos\left[\frac{(-2eV+2\mu_N) R_{\delta,\epsilon}}{\hbar v_F} \right]
    \cos\left[\frac{(-2eV+2\mu_N) R_{\beta,\epsilon}}{\hbar v_F} \right]
    \cos\left[\frac{(-2eV+2\mu_N) R_{\gamma,\beta}}{\hbar v_F} \right]
    .
\end{eqnarray}

\subsubsection{Final expression for the $\cos \Psi_Q$-component of the ${Q}'$-quartet supercurrent}
\label{subsubtata3}

In this subsubsection of the SM, we conclude the calculation of the
$Q'$-quartet supercurrent. The derivative of the ${Q}'$-quartet
supercurrent with respect to the electrochemical potential $\mu_N$ is
the following:
\begin{eqnarray}
  \label{eq:result5}
    && \frac{\partial I_{{Q}'}}{\partial \mu_N} =
  \frac{\left(\Sigma_L\right)^4 \left(\Sigma_R\right)^4
    \left(\Sigma_B\right)^8} {2 W^{16}} \times
  \cos\left(\Psi_Q\right) \times \frac{1}{(k_F
    R_{\alpha,\delta})(k_F R_{\delta,\epsilon}) (k_F
    R_{\beta,\epsilon}) (k_F R_{\gamma,\beta})} \times\\ \nonumber && \left\{
  \cos\left[\frac{(-2eV+2\mu_N) R_{\alpha,\delta}}{\hbar v_F} \right]
  \cos\left[\frac{(-2eV+2\mu_N) R_{\delta,\epsilon}}{\hbar v_F}
    \right] \cos\left[\frac{(-2eV+2\mu_N) R_{\beta,\epsilon}}{\hbar
      v_F} \right] \cos\left[\frac{(-2eV+2\mu_N)
      R_{\gamma,\beta}}{\hbar v_F} \right] \right.\\ \nonumber &&+
  \cos\left[\frac{(2eV+2\mu_N) R_{\alpha,\delta}}{\hbar v_F} \right]
  \cos\left[\frac{(2eV+2\mu_N) R_{\delta,\epsilon}}{\hbar v_F} \right]
  \cos\left[\frac{(2eV+2\mu_N) R_{\beta,\epsilon}}{\hbar v_F} \right]
  \cos\left[\frac{(2eV+2\mu_N) R_{\gamma,\beta}}{\hbar v_F}
    \right]\\ \nonumber &&+ \cos\left[\frac{(-6eV+2\mu_N)
      R_{\alpha,\delta}}{\hbar v_F} \right]
  \cos\left[\frac{(-6eV+2\mu_N) R_{\delta,\epsilon}}{\hbar v_F}
    \right] \cos\left[\frac{(-6eV+2\mu_N) R_{\beta,\epsilon}}{\hbar
      v_F} \right] \cos\left[\frac{(-6eV+2\mu_N)
      R_{\gamma,\beta}}{\hbar v_F} \right]\\ \nonumber &&+\left.
  \cos\left[\frac{(6eV+2\mu_N) R_{\alpha,\delta}}{\hbar v_F} \right]
  \cos\left[\frac{(6eV+2\mu_N) R_{\delta,\epsilon}}{\hbar v_F} \right]
  \cos\left[\frac{(6eV+2\mu_N) R_{\beta,\epsilon}}{\hbar v_F} \right]
  \cos\left[\frac{(6eV+2\mu_N) R_{\gamma,\beta}}{\hbar v_F} \right]
  \right\}
  .
\end{eqnarray}
{The dimensionless $Q'$-quartet current susceptibility
  takes the following form:
\begin{eqnarray}
  \label{eq:i'Q'}
    && \frac{\partial i'_{{Q}'}(eV,\mu_N)}{\partial \mu_N} =
  \frac{1}{(k_F
    R_{\alpha,\delta})(k_F R_{\delta,\epsilon}) (k_F
    R_{\beta,\epsilon}) (k_F R_{\gamma,\beta})} \times\\ \nonumber && \left\{
  \cos\left[\frac{(-2eV+2\mu_N) R_{\alpha,\delta}}{\hbar v_F} \right]
  \cos\left[\frac{(-2eV+2\mu_N) R_{\delta,\epsilon}}{\hbar v_F}
    \right] \cos\left[\frac{(-2eV+2\mu_N) R_{\beta,\epsilon}}{\hbar
      v_F} \right] \cos\left[\frac{(-2eV+2\mu_N)
      R_{\gamma,\beta}}{\hbar v_F} \right] \right.\\ \nonumber &&+
  \cos\left[\frac{(2eV+2\mu_N) R_{\alpha,\delta}}{\hbar v_F} \right]
  \cos\left[\frac{(2eV+2\mu_N) R_{\delta,\epsilon}}{\hbar v_F} \right]
  \cos\left[\frac{(2eV+2\mu_N) R_{\beta,\epsilon}}{\hbar v_F} \right]
  \cos\left[\frac{(2eV+2\mu_N) R_{\gamma,\beta}}{\hbar v_F}
    \right]\\ \nonumber &&+ \cos\left[\frac{(-6eV+2\mu_N)
      R_{\alpha,\delta}}{\hbar v_F} \right]
  \cos\left[\frac{(-6eV+2\mu_N) R_{\delta,\epsilon}}{\hbar v_F}
    \right] \cos\left[\frac{(-6eV+2\mu_N) R_{\beta,\epsilon}}{\hbar
      v_F} \right] \cos\left[\frac{(-6eV+2\mu_N)
      R_{\gamma,\beta}}{\hbar v_F} \right]\\ \nonumber &&+\left.
  \cos\left[\frac{(6eV+2\mu_N) R_{\alpha,\delta}}{\hbar v_F} \right]
  \cos\left[\frac{(6eV+2\mu_N) R_{\delta,\epsilon}}{\hbar v_F} \right]
  \cos\left[\frac{(6eV+2\mu_N) R_{\beta,\epsilon}}{\hbar v_F} \right]
  \cos\left[\frac{(6eV+2\mu_N) R_{\gamma,\beta}}{\hbar v_F} \right]
  \right\}
  .
\end{eqnarray}
Eq.~(5) in the main text relates $\partial i'_{Q'}(eV,\mu_N)/\partial
\mu_N$ in Eq.~(\ref{eq:i'Q'}) to $i_{Q'}(\omega)$ in
Eq.~(\ref{eq:iQ'-omega}).}

\section{Calculation of the Fano factor}
\label{sec:Fano}

In this section of the SM, we demonstrate the value $2N$ of the Fano
factor associated to the Floquet-MAR processes described by the snake
diagrams of order-$N$, see Eq.~(6) in the main text. General arguments
are presented in subsection~\ref{sec:general-F} and a complementary
demonstration is provided in subsection~\ref{sec:vi}.

\subsection{General arguments}
\label{sec:general-F}
In this subsection of the SM, we present general arguments about the
quantum noise of the $Q$-quartets, $O$-octets, $Q'$-quartets and
higher-order multipairs.  We start with an Andreev interferometer of
order-$N=1$, and the resulting current $I_L$ and quantum noise
cross-correlations $S_{L,R}$ are such that
\begin{eqnarray}
  \frac{\partial I_L}{\partial \mu_N}&=& A_1 \cos\Psi
  \cos\left(\frac{2\omega
    R_{\alpha,\beta}}{v_F}\right)\\ \frac{\partial S_{L,R}}{\partial
    \mu_N}&=& B_1 \cos\Psi \cos\left(\frac{2\omega
    R_{\alpha,\beta}}{v_F}\right) .
\end{eqnarray}
We now evaluate the ratio $B_N/A_N$ for a snake diagram of
order-$N$. Inspecting the corresponding diagrams leads to the
following prefactors:

(i) There are two terms in the current, see the above
Eq.~(\ref{eq:A1}) in terms of $\hat{\Sigma} \hat{G}^{+,-}-\hat{\Sigma}
\hat{G}^{+,-}$ and four terms in the current-current
cross-correlations, see the above Eq.~(\ref{eq:K}).

(ii) Expanding over the Nambu labels doubles the number of terms, both
in the current and in the current-current cross-correlations.

(iii) The bare Keldysh Green's function $\hat{g}^{+,-}$ is expressed
as the product between the distribution functions and the difference
between the advanced and the retarded Green's functions, thus
producing two terms. Similarly, the product
$\hat{g}^{+,-}\hat{g}^{-,+}$ produces four terms in the
current-current cross-correlations.

(iv) Averaging the modes at the scale of the Fermi wave-length
produces a factor of $(1/2)^N$, both in the current and in the
current-current cross-correlations.

(v) For the current, there are $2N$ positions of $\hat{g}^{+,-}$ and,
in the noise, there are $N^2$ positions of the $\hat{g}^{+,-}$ and the
$\hat{g}^{-,+}$ bare Keldysh Green's functions.

(vi) Including the $\tau \leftrightarrow \tau'$ term in
Eq.~(\ref{eq:K}) produces a factor of two in the current-current
cross-correlations.

(vii) Differentiating with respect to $\mu_N$ produces a factor of $1/2$
in the current-current cross-correlations. This statement is demonstrated
in the forthcoming subsection~\ref{sec:vi}.

We deduce that, overall, the Fano factor takes the value $B_N/A_N=2N$,
see Eq.~(6) in the main text.

\subsection{Demonstration of statement (vii)}

\label{sec:vi}

In this subsection of the SM, we demonstrate the above item
(vii). Specifically, we differentiate the following expression
\begin{equation}
  X=\theta\left(\epsilon\mu_N + p e V - \omega\right)
  \theta\left(\omega+\epsilon'\mu_N+p'eV\right)
\end{equation}
with respect to $\mu_N$, where $\epsilon,\epsilon'=\pm$ and $p$, $p'$
are two positive or negative integers that label the Floquet
replica. We obtain
\begin{equation}
  \frac{\partial X}{\partial \mu_N}=
  \theta\left[(\epsilon+\epsilon')\mu_N+(p+p')eV\right]
  \left\{ \epsilon \delta\left(\epsilon\mu_N+peV-\omega\right)
  +\epsilon' \delta\left(\epsilon'\mu_N+p'eV-\omega\right)\right\}
  .
\end{equation}
\begin{figure*}[htb]
  \includegraphics[width=.49\textwidth]{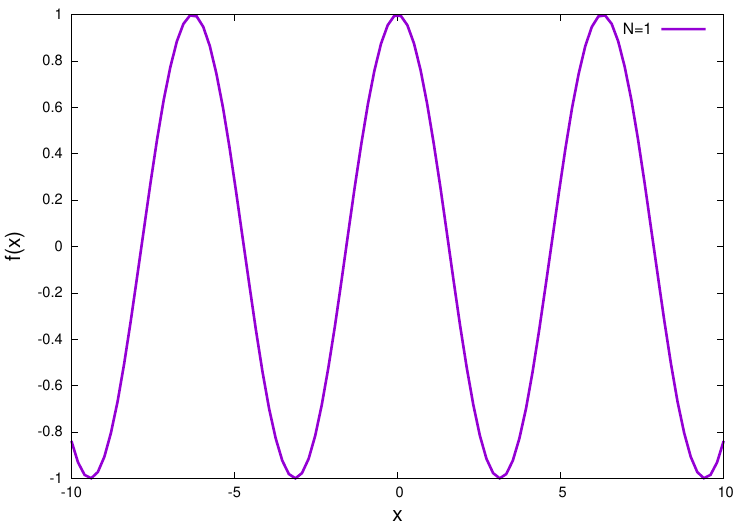}\includegraphics[width=.49\textwidth]{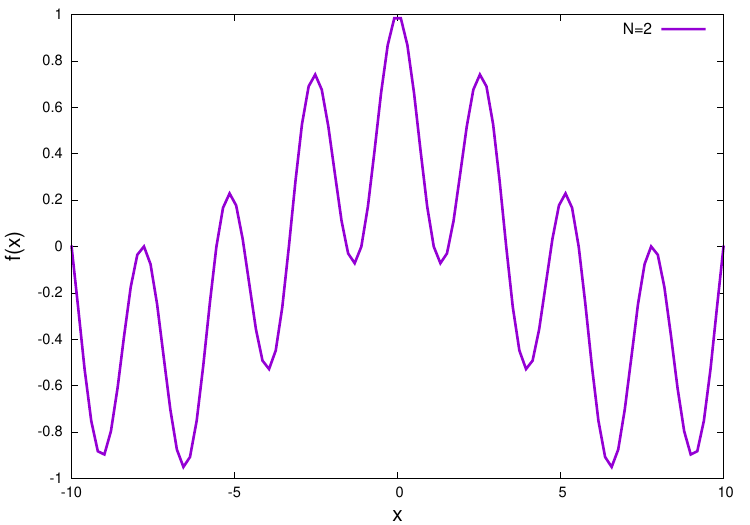}

  \includegraphics[width=.49\textwidth]{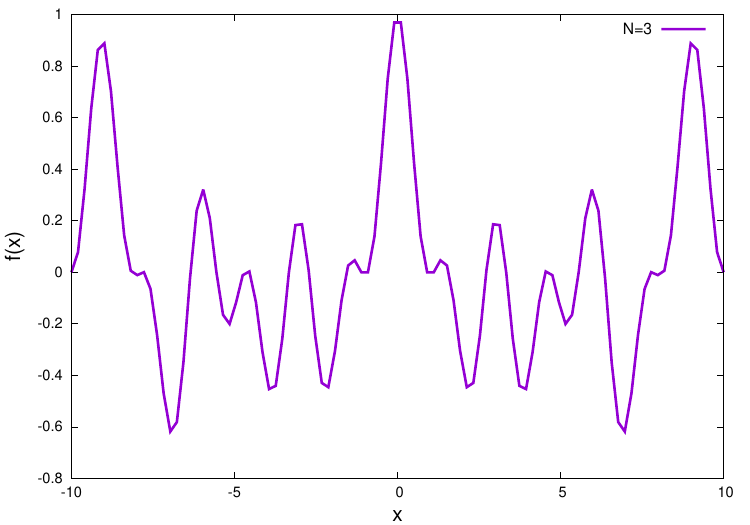}\includegraphics[width=.49\textwidth]{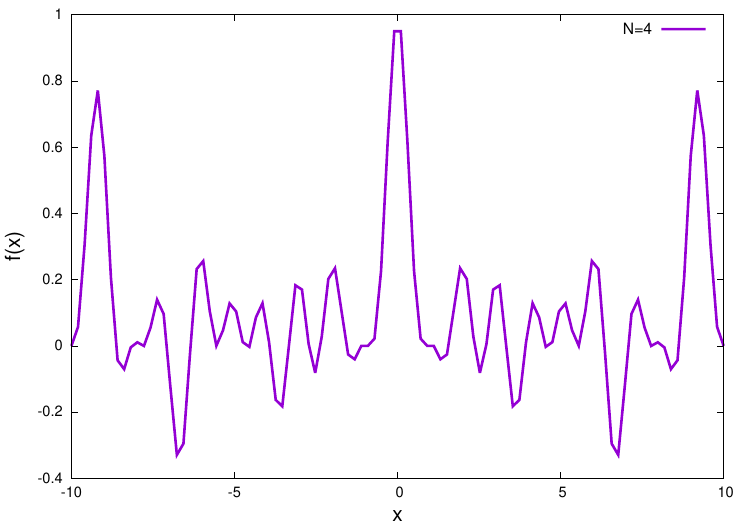}

  \caption{The figure shows the emergence of an anomaly at $x=0$, as
    $N$ increases in Eq.~(\ref{eq:N}).
    \label{fig:N}}
\end{figure*}
Exchanging the ``1'' and the ``2'' Nambu labels amounts to changing
$\epsilon$ into $-\epsilon$ and $\epsilon'$ into $-\epsilon'$, and
changing $p$ into $-p$ and $p'$ into $-p'$. Then, $\partial X/\partial \mu_N$
is changed into
\begin{equation}
  -\theta\left[-(\epsilon+\epsilon')\mu_N-(p+p')eV\right]
  \left\{\epsilon \delta\left(\epsilon\mu_N+peV-\omega\right)
  +\epsilon'\delta\left(\epsilon'\mu_N+p'eV-\omega\right) \right\}
  .
\end{equation}
We deduce that taking the $\partial/\partial \mu_N$ derivative makes
cancel half of the terms.

\begin{figure*}[htb]
  \includegraphics[width=.49\textwidth]{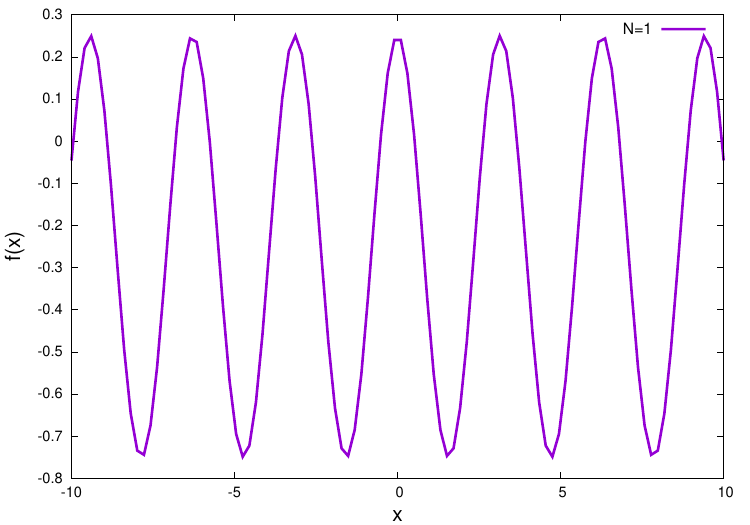}\includegraphics[width=.49\textwidth]{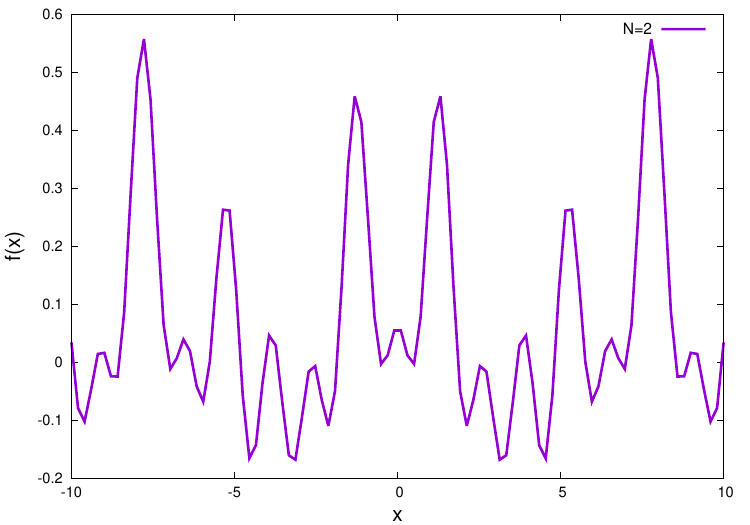}

  \includegraphics[width=.49\textwidth]{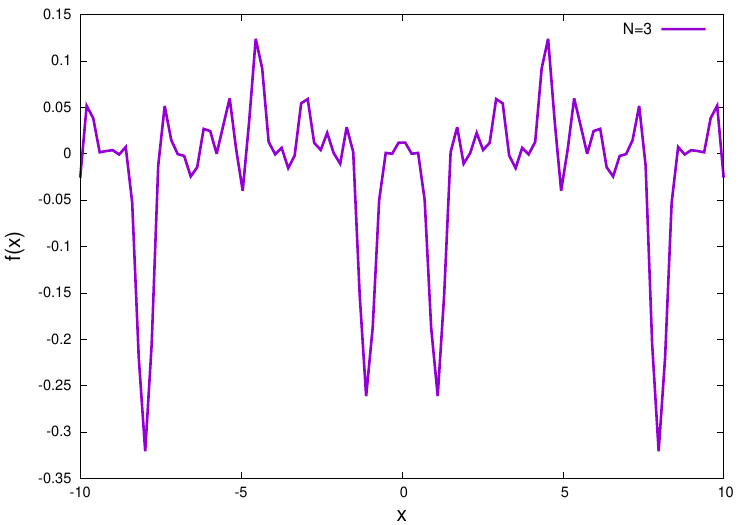}\includegraphics[width=.49\textwidth]{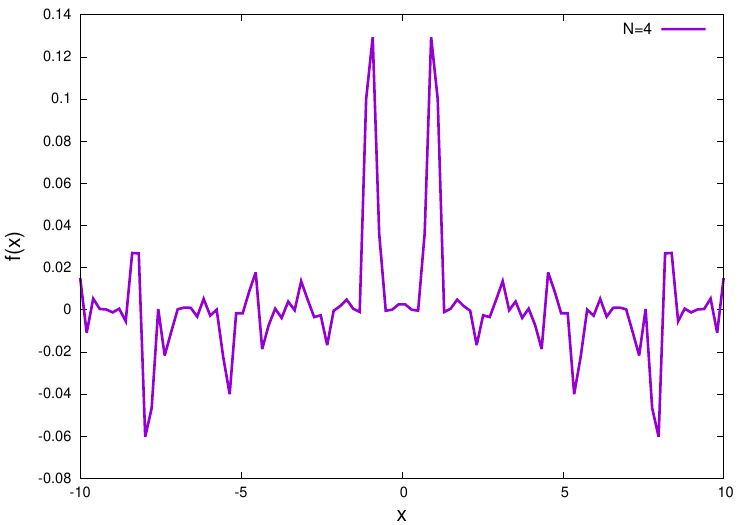}
  \caption{The figure shows the absence of an anomaly at $x=0$ as $N$
    increases in Eq.~(\ref{eq:N2}) with $\theta=\pi/3$.
    \label{fig:N2}}
\end{figure*}

\section{Toy-model for the emergence of Floquet resonances in the current susceptibility}
\label{sec:toy}

In this subsubsection, we present simple arguments inspired by the above full
analytical formula. To illustrate the emergence of finite-bias resonances in the
current and noise, we evaluate the following
toy-model:
\begin{equation}
  \label{eq:N}
  f(x)=\prod_{k=1}^N \cos(\sqrt{k} x)
  ,
\end{equation}
where the product of the $\cos$ is a simplification of the product of
the modes in the snake diagrams, and the $\sqrt k$ are generally
incommensurate.  The function $f(x)$ given by Eq.~(\ref{eq:N}) becomes
more peaked around zero as $N$ increases, as shown in Fig.~\ref{fig:N}
of the SM. This toy-model explains why resonances emerge in Figure 3
(in the main text).

Conversely, the ballistic case can be simulated from Eq.~(\ref{eq:g-g-ball}), using
the following toy-model:
\begin{equation}
  \label{eq:N2}
  f(x)=\prod_{k=1}^N \left[\cos\theta \cos(\sqrt{k} x)-\sin \theta\sin(\sqrt{k} x)\right]
  \times \left[\cos\theta \cos(\sqrt{k} x)+\sin \theta\sin(\sqrt{k} x)\right]
  ,
\end{equation}
where the $\theta$-term in Eq.~(\ref{eq:N2}) reflects the
$k_F$-contribution in Eq.~(\ref{eq:g-g-ball}).  Fig.~\ref{fig:N2} of
the SM reveals the resulting absence of $x=0$ anomaly at finite angle
$\theta$.